\documentclass[longauth]{aa}
\usepackage{graphicx}
\usepackage{subfigure}
\usepackage{txfonts}
\usepackage[range-units=single,range-phrase=\text{--}]{siunitx}
\usepackage{xspace}
\usepackage{makecell}
\usepackage[switch]{lineno}
\usepackage[colorlinks,allcolors=blue]{hyperref}
\newcommand{\gam}{$\gamma$\xspace}
\newcommand{\gammapy}{\textsc{Gammapy}\xspace}
\newcommand{\fermipy}{\textsc{Fermipy}\xspace}
\newcommand{\hess}{H.E.S.S.\xspace}
\newcommand{\fermilat}{\emph{Fermi}-LAT\xspace}
\newcommand{\hessj}{HESS~J1809$-$193\xspace}
\newcommand{\psrj}{PSR~J1809$-$1917\xspace}
\newcommand{\psrjother}{PSR~J1811$-$1925\xspace}
\newcommand{\snrN}{G011.1+00.1\xspace}
\newcommand{\snrS}{G011.0$-$00.0\xspace}
\newcommand{\fermijten}{J1810.3$-$1925e\xspace}
\newcommand{\fermijeleven}{J1811.5$-$1925\xspace}

\DeclareSIUnit\pc{pc}
\DeclareSIUnit\kpc{kpc}
\DeclareSIUnit\erg{erg}
\DeclareSIUnit\year{yr}
\DeclareSIUnit\TeV{TeV}
\DeclareSIUnit\GeV{GeV}
\DeclareSIUnit\PeV{PeV}
\DeclareSIUnit\gauss{G}
\DeclareSIUnit\mas{mas}

\makeatletter
\renewcommand*\aa@pageof{, page \thepage{} of \pageref*{LastPage}}
\makeatother

\begin{document}


\title{\hessj: a halo of escaped electrons\\around a pulsar wind nebula?}
\titlerunning{\hessj: a halo of escaped electrons around a pulsar wind nebula?}

\author{
\begin{small}
F.~Aharonian \inst{\ref{DIAS},\ref{MPIK}}
\and F.~Ait~Benkhali \inst{\ref{LSW}}
\and J.~Aschersleben \inst{\ref{Groningen}}
\and H.~Ashkar \inst{\ref{LLR}}
\and M.~Backes \inst{\ref{UNAM},\ref{NWU}}
\and V.~Barbosa~Martins \inst{\ref{DESY}}
\and R.~Batzofin \inst{\ref{Wits}}
\and Y.~Becherini \inst{\ref{APC},\ref{Linnaeus}}
\and D.~Berge \inst{\ref{DESY},\ref{HUB}}
\and M.~B\"ottcher \inst{\ref{NWU}}
\and C.~Boisson \inst{\ref{LUTH}}
\and J.~Bolmont \inst{\ref{LPNHE}}
\and J.~Borowska \inst{\ref{HUB}}
\and M.~Bouyahiaoui \inst{\ref{MPIK}}
\and F.~Bradascio \inst{\ref{CEA}}
\and M.~Breuhaus \inst{\ref{MPIK}}
\and R.~Brose \inst{\ref{DIAS}}
\and F.~Brun \inst{\ref{CEA}}
\and B.~Bruno \inst{\ref{ECAP}}
\and T.~Bulik \inst{\ref{UWarsaw}}
\and C.~Burger-Scheidlin \inst{\ref{DIAS}}
\and T.~Bylund \inst{\ref{Linnaeus}}
\and S.~Caroff \inst{\ref{LAPP}}
\and S.~Casanova \inst{\ref{IFJPAN}}
\and J.~Celic \inst{\ref{ECAP}}
\and M.~Cerruti \inst{\ref{APC}}
\and P.~Chambery \inst{\ref{CENBG}}
\and T.~Chand \inst{\ref{NWU}}
\and A.~Chen \inst{\ref{Wits}}
\and J.~Chibueze \inst{\ref{NWU}}
\and O.~Chibueze \inst{\ref{NWU}}
\and J.~Damascene~Mbarubucyeye \inst{\ref{DESY}}
\and A.~Djannati-Ata\"i \inst{\ref{APC}}
\and A.~Dmytriiev \inst{\ref{NWU}}
\and S.~Einecke \inst{\ref{Adelaide}}
\and J.-P.~Ernenwein \inst{\ref{CPPM}}
\and K.~Feijen \inst{\ref{Adelaide}}
\and M.~Filipovic \inst{\ref{Sydney}}
\and G.~Fontaine \inst{\ref{LLR}}
\and M.~F\"u{\ss}ling \inst{\ref{DESY}}
\and S.~Funk \inst{\ref{ECAP}}
\and S.~Gabici \inst{\ref{APC}}
\and Y.A.~Gallant \inst{\ref{LUPM}}
\and S.~Ghafourizadeh \inst{\ref{LSW}}
\and G.~Giavitto \inst{\ref{DESY}}
\and L.~Giunti \inst{\ref{APC},\ref{CEA}}
\and D.~Glawion \inst{\ref{ECAP}}
\and P.~Goswami \inst{\ref{NWU}}
\and G.~Grolleron \inst{\ref{LPNHE}}
\and M.-H.~Grondin \inst{\ref{CENBG}}
\and L.~Haerer \inst{\ref{MPIK}}
\and J.A.~Hinton \inst{\ref{MPIK}}
\and W.~Hofmann \inst{\ref{MPIK}}
\and T.~L.~Holch \inst{\ref{DESY}}
\and M.~Holler \inst{\ref{Innsbruck}}
\and D.~Horns \inst{\ref{UHH}}
\and Zhiqiu~Huang \inst{\ref{MPIK}}
\and M.~Jamrozy \inst{\ref{UJK}}
\and F.~Jankowsky \inst{\ref{LSW}}
\and V.~Joshi \inst{\ref{ECAP}}$^{,}$\footnotemark[1] 
\and I.~Jung-Richardt \inst{\ref{ECAP}}
\and E.~Kasai \inst{\ref{UNAM}}
\and K.~Katarzy{\'n}ski \inst{\ref{NCUT}}
\and B.~Kh\'elifi \inst{\ref{APC}}
\and W.~Klu\'{z}niak \inst{\ref{NCAC}}
\and Nu.~Komin \inst{\ref{Wits}}
\and K.~Kosack \inst{\ref{CEA}}
\and D.~Kostunin \inst{\ref{DESY}}
\and R.G.~Lang \inst{\ref{ECAP}}
\and S.~Le~Stum \inst{\ref{CPPM}}
\and F.~Leitl \inst{\ref{ECAP}}
\and A.~Lemi\`ere \inst{\ref{APC}}
\and M.~Lemoine-Goumard \inst{\ref{CENBG}}
\and J.-P.~Lenain \inst{\ref{LPNHE}}
\and F.~Leuschner \inst{\ref{IAAT}}
\and T.~Lohse \inst{\ref{HUB}}
\and A.~Luashvili \inst{\ref{LUTH}}
\and I.~Lypova \inst{\ref{LSW}}
\and J.~Mackey \inst{\ref{DIAS}}
\and D.~Malyshev \inst{\ref{IAAT}}
\and D.~Malyshev \inst{\ref{ECAP}}
\and V.~Marandon \inst{\ref{MPIK}}
\and P.~Marchegiani \inst{\ref{Wits}}
\and A.~Marcowith \inst{\ref{LUPM}}
\and P.~Marinos \inst{\ref{Adelaide}}
\and G.~Mart\'i-Devesa \inst{\ref{Innsbruck}}
\and R.~Marx \inst{\ref{LSW}}
\and A.~Mitchell \inst{\ref{ECAP}}
\and R.~Moderski \inst{\ref{NCAC}}
\and L.~Mohrmann \inst{\ref{MPIK}}$^{,}$\thanks{Corresponding authors;\\\email{\href{mailto:contact.hess@hess-experiment.eu}{contact.hess@hess-experiment.eu}}}
\and A.~Montanari \inst{\ref{CEA}}
\and E.~Moulin \inst{\ref{CEA}}
\and J.~Muller \inst{\ref{LLR}}
\and K.~Nakashima \inst{\ref{ECAP}}
\and M.~de~Naurois \inst{\ref{LLR}}
\and J.~Niemiec \inst{\ref{IFJPAN}}
\and A.~Priyana~Noel \inst{\ref{UJK}}
\and S.~Ohm \inst{\ref{DESY}}
\and L.~Olivera-Nieto \inst{\ref{MPIK}}
\and E.~de~Ona~Wilhelmi \inst{\ref{DESY}}
\and M.~Ostrowski \inst{\ref{UJK}}
\and S.~Panny \inst{\ref{Innsbruck}}
\and M.~Panter \inst{\ref{MPIK}}
\and R.D.~Parsons \inst{\ref{HUB}}
\and D.A.~Prokhorov \inst{\ref{Amsterdam}}
\and G.~P\"uhlhofer \inst{\ref{IAAT}}
\and M.~Punch \inst{\ref{APC}}
\and A.~Quirrenbach \inst{\ref{LSW}}
\and P.~Reichherzer \inst{\ref{CEA}}
\and A.~Reimer \inst{\ref{Innsbruck}}
\and O.~Reimer \inst{\ref{Innsbruck}}
\and M.~Renaud \inst{\ref{LUPM}}
\and B.~Reville \inst{\ref{MPIK}}
\and F.~Rieger \inst{\ref{MPIK}}
\and G.~Rowell \inst{\ref{Adelaide}}
\and B.~Rudak \inst{\ref{NCAC}}
\and V.~Sahakian \inst{\ref{Yerevan}}
\and A.~Santangelo \inst{\ref{IAAT}}
\and M.~Sasaki \inst{\ref{ECAP}}
\and H.M.~Schutte \inst{\ref{NWU}}
\and U.~Schwanke \inst{\ref{HUB}}
\and J.N.S.~Shapopi \inst{\ref{UNAM}}
\and H.~Sol \inst{\ref{LUTH}}
\and A.~Specovius \inst{\ref{ECAP}}
\and S.~Spencer \inst{\ref{ECAP}}
\and {\L.}~Stawarz \inst{\ref{UJK}}
\and R.~Steenkamp \inst{\ref{UNAM}}
\and S.~Steinmassl \inst{\ref{MPIK}}
\and I.~Sushch \inst{\ref{NWU}}
\and H.~Suzuki \inst{\ref{Konan}}
\and T.~Takahashi \inst{\ref{KAVLI}}
\and T.~Tanaka \inst{\ref{Konan}}
\and R.~Terrier \inst{\ref{APC}}
\and C.~Thorpe-Morgan \inst{\ref{IAAT}}
\and M.~Tsirou \inst{\ref{MPIK}}
\and N.~Tsuji \inst{\ref{RIKKEN}}
\and Y.~Uchiyama \inst{\ref{Rikkyo}}
\and C.~van~Eldik \inst{\ref{ECAP}}
\and M.~Vecchi \inst{\ref{Groningen}}
\and J.~Veh \inst{\ref{ECAP}}
\and C.~Venter \inst{\ref{NWU}}
\and J.~Vink \inst{\ref{Amsterdam}}
\and T.~Wach \inst{\ref{ECAP}}
\and S.J.~Wagner \inst{\ref{LSW}}
\and R.~White \inst{\ref{MPIK}}
\and A.~Wierzcholska \inst{\ref{IFJPAN}}
\and Yu~Wun~Wong \inst{\ref{ECAP}}
\and M.~Zacharias \inst{\ref{LSW},\ref{NWU}}
\and D.~Zargaryan \inst{\ref{DIAS}}
\and A.A.~Zdziarski \inst{\ref{NCAC}}
\and A.~Zech \inst{\ref{LUTH}}
\and S.~Zouari \inst{\ref{APC}}
\and N.~\.Zywucka \inst{\ref{NWU}}
(\hess Collaboration)
\end{small}
}
\authorrunning{F. Aharonian et al.}

\institute{
Dublin Institute for Advanced Studies, 31 Fitzwilliam Place, Dublin 2, Ireland \label{DIAS} \and
Max-Planck-Institut f\"ur Kernphysik, P.O. Box 103980, D 69029 Heidelberg, Germany \label{MPIK} \and
Landessternwarte, Universit\"at Heidelberg, K\"onigstuhl, D 69117 Heidelberg, Germany \label{LSW} \and
Kapteyn Astronomical Institute, University of Groningen, Landleven 12, 9747 AD Groningen, The Netherlands \label{Groningen} \and
Laboratoire Leprince-Ringuet, École Polytechnique, CNRS, Institut Polytechnique de Paris, F-91128 Palaiseau, France \label{LLR} \and
University of Namibia, Department of Physics, Private Bag 13301, Windhoek 10005, Namibia \label{UNAM} \and
Centre for Space Research, North-West University, Potchefstroom 2520, South Africa \label{NWU} \and
DESY, D-15738 Zeuthen, Germany \label{DESY} \and
School of Physics, University of the Witwatersrand, 1 Jan Smuts Avenue, Braamfontein, Johannesburg, 2050 South Africa \label{Wits} \and
Université de Paris, CNRS, Astroparticule et Cosmologie, F-75013 Paris, France \label{APC} \and
Department of Physics and Electrical Engineering, Linnaeus University,  351 95 V\"axj\"o, Sweden \label{Linnaeus} \and
Institut f\"ur Physik, Humboldt-Universit\"at zu Berlin, Newtonstr. 15, D 12489 Berlin, Germany \label{HUB} \and
Laboratoire Univers et Théories, Observatoire de Paris, Université PSL, CNRS, Université de Paris, 92190 Meudon, France \label{LUTH} \and
Sorbonne Universit\'e, Universit\'e Paris Diderot, Sorbonne Paris Cit\'e, CNRS/IN2P3, Laboratoire de Physique Nucl\'eaire et de Hautes Energies, LPNHE, 4 Place Jussieu, F-75252 Paris, France \label{LPNHE} \and
IRFU, CEA, Universit\'e Paris-Saclay, F-91191 Gif-sur-Yvette, France \label{CEA} \and
Friedrich-Alexander-Universit\"at Erlangen-N\"urnberg, Erlangen Centre for Astroparticle Physics, Erwin-Rommel-Str. 1, D 91058 Erlangen, Germany \label{ECAP} \and
Astronomical Observatory, The University of Warsaw, Al. Ujazdowskie 4, 00-478 Warsaw, Poland \label{UWarsaw} \and
Université Savoie Mont Blanc, CNRS, Laboratoire d'Annecy de Physique des Particules - IN2P3, 74000 Annecy, France \label{LAPP} \and
Instytut Fizyki J\c{a}drowej PAN, ul. Radzikowskiego 152, 31-342 Krak{\'o}w, Poland \label{IFJPAN} \and
Universit\'e Bordeaux, CNRS, LP2I Bordeaux, UMR 5797, F-33170 Gradignan, France \label{CENBG} \and
School of Physical Sciences, University of Adelaide, Adelaide 5005, Australia \label{Adelaide} \and
Aix Marseille Universit\'e, CNRS/IN2P3, CPPM, Marseille, France \label{CPPM} \and
School of Science, Western Sydney University, Locked Bag 1797, Penrith South DC, NSW 2751, Australia \label{Sydney} \and
Laboratoire Univers et Particules de Montpellier, Universit\'e Montpellier, CNRS/IN2P3,  CC 72, Place Eug\`ene Bataillon, F-34095 Montpellier Cedex 5, France \label{LUPM} \and
Institut f\"ur Astro- und Teilchenphysik, Leopold-Franzens-Universit\"at Innsbruck, A-6020 Innsbruck, Austria \label{Innsbruck} \and
Universit\"at Hamburg, Institut f\"ur Experimentalphysik, Luruper Chaussee 149, D 22761 Hamburg, Germany \label{UHH} \and
Obserwatorium Astronomiczne, Uniwersytet Jagiello{\'n}ski, ul. Orla 171, 30-244 Krak{\'o}w, Poland \label{UJK} \and
Institute of Astronomy, Faculty of Physics, Astronomy and Informatics, Nicolaus Copernicus University,  Grudziadzka 5, 87-100 Torun, Poland \label{NCUT} \and
Nicolaus Copernicus Astronomical Center, Polish Academy of Sciences, ul. Bartycka 18, 00-716 Warsaw, Poland \label{NCAC} \and
Institut f\"ur Astronomie und Astrophysik, Universit\"at T\"ubingen, Sand 1, D 72076 T\"ubingen, Germany \label{IAAT} \and
GRAPPA, Anton Pannekoek Institute for Astronomy, University of Amsterdam,  Science Park 904, 1098 XH Amsterdam, The Netherlands \label{Amsterdam} \and
Yerevan Physics Institute, 2 Alikhanian Brothers St., 375036 Yerevan, Armenia \label{Yerevan} \and
Department of Physics, Konan University, 8-9-1 Okamoto, Higashinada, Kobe, Hyogo 658-8501, Japan \label{Konan} \and
Kavli Institute for the Physics and Mathematics of the Universe (WPI), The University of Tokyo Institutes for Advanced Study (UTIAS), The University of Tokyo, 5-1-5 Kashiwa-no-Ha, Kashiwa, Chiba, 277-8583, Japan \label{KAVLI} \and
RIKEN, 2-1 Hirosawa, Wako, Saitama 351-0198, Japan \label{RIKKEN} \and
Department of Physics, Rikkyo University, 3-34-1 Nishi-Ikebukuro, Toshima-ku, Tokyo 171-8501, Japan \label{Rikkyo}
}

\date{\today}

\abstract
{
  \hessj is an unassociated very-high-energy \gam-ray source located on the Galactic plane.
  While it has been connected to the nebula of the energetic pulsar \psrj, supernova remnants and molecular clouds present in the vicinity also constitute possible associations.
  Recently, the detection of \gam-ray emission up to energies of $\sim$\SI{100}{\TeV} with the HAWC observatory has led to renewed interest in \hessj.
}
{
  We aim to understand the origin of the \gam-ray emission of \hessj.
}
{
  We analysed \SI{93.2}{\hour} of data taken on \hessj above \SI{0.27}{\TeV} with the High Energy Stereoscopic System (\hess), using a multi-component, three-dimensional likelihood analysis.
  In addition, we provide a new analysis of \SI{12.5}{\year} of \fermilat data above \SI{1}{\GeV} within the region of \hessj.
  The obtained results are interpreted in a time-dependent modelling framework.
}
{
  For the first time, we were able to resolve the emission detected with \hess into two components:
  an extended component (modelled as an elongated Gaussian with a 1-$\sigma$ semi-major~/ semi-minor axis of $\sim$0.62$^\circ$ / $\sim$0.35$^\circ$) that exhibits a spectral cut-off at $\sim$\SI{13}{TeV}, and a compact component (modelled as a symmetric Gaussian with a 1-$\sigma$ radius of $\sim$0.1$^\circ$) that is located close to \psrj and shows no clear spectral cut-off.
  The \fermilat analysis also revealed extended \gam-ray emission, on scales similar to that of the extended \hess component.
}
{
  Our modelling indicates that based on its spectrum and spatial extent, the extended \hess component is likely caused by inverse Compton emission from old electrons that form a halo around the pulsar wind nebula.
  The compact component could be connected to either the pulsar wind nebula or the supernova remnant and molecular clouds.
  Due to its comparatively steep spectrum, modelling the \fermilat emission together with the \hess components is not straightforward.
} 


\maketitle

\section{Introduction}

The \hess Galactic Plane Survey \citep[HGPS;][]{HESS_HGPS_2018} has revealed a large number of Galactic \gam-ray sources in the very-high-energy (VHE; $E>\SI{100}{\GeV}$) domain.
While a number of these sources could be firmly associated with multi-wavelength counterparts, a large fraction of the sources remain without firm association.
Among the firmly identified sources, the vast majority are either pulsar wind nebulae (PWNe) or supernova remnants (SNRs), or composite systems.
In this paper, we study the unassociated source \hessj, which has been discovered using early HGPS observations \citep{HESS_J1809_2007}.

Identifying the physical counterpart of \hessj is particularly challenging due to the presence of several plausible associations in its vicinity.
For instance, the region harbours two energetic pulsars: PSR~J1811$-$1925 ($\dot{E}=\SI{6.4e36}{\erg\per\second}$, $d\sim \SI{5}{\kpc}$) and, noteworthy in particular, \psrj ($\dot{E}=\SI{1.8e36}{\erg\per\second}$, $d\sim \SI{3.3}{\kpc}$) \citep{Manchester2005}, which powers an X-ray PWN with an extension of $\sim$3\arcmin{} \citep{Kargaltsev2007,Anada2010,Klingler2018,Klingler2020}.
Located nearby is also a transient X-ray magnetar, XTE~J1810$-$197 \citep{Alford2016}.
On the other hand, \hessj is also spatially coincident with several SNRs, most notably \snrN and \snrS \citep{Green2019}, as well as with molecular clouds \citep{Castelletti2016,Voisin2019}.
This leaves open the possibility to interpret the \gam-ray emission as originating from high-energy electrons\footnote{In this paper, we use the term `electrons' to refer to both electrons and positrons.} -- most likely provided by one of the pulsars -- that up-scatter photons from ambient radiation fields to \gam-ray energies via the Inverse Compton (IC) process (`leptonic scenario'), or as being due to interactions of high-energy cosmic-ray nuclei -- accelerated, for example, at SNR shock fronts -- within nearby molecular clouds (`hadronic scenario').
In the \hessj discovery paper \citep{HESS_J1809_2007}, as well as in two follow-up studies presented shortly afterwards \citep{Komin2007,Renaud2008}, the authors have shown that the PWN surrounding \psrj can naturally explain the observed \gam-ray emission in a leptonic scenario, and thus represents the most likely association.
However, \citet{Castelletti2016} and \citet{Araya2018} have subsequently put forward an interpretation in a hadronic scenario involving the SNRs and molecular clouds found within the region.

Recently, the detection of \gam-ray emission from \hessj with the High Altitude Water Cherenkov Observatory (HAWC) above \SI{56}{\TeV} (and quite possibly above \SI{100}{\TeV}; \citeauthor{HAWC2020} \citeyear{HAWC2020}; see also \citeauthor{Goodman2022} \citeyear{Goodman2022}) has added to the motivation to identify its origin.
In a hadronic scenario, this detection would make \hessj a good `PeVatron' candidate -- that is, a source capable of accelerating cosmic-ray nuclei to PeV energies.
The identification of such PeVatrons is regarded as decisive in the quest for unveiling the origin of Galactic cosmic rays \citep[e.g.][]{Berezinskii1990,Aharonian2013,Cristofari2021}.
On the other hand, should the leptonic interpretation hold, the detection by HAWC would demonstrate that \hessj is a fascinating laboratory for the study of high-energy electrons and their propagation -- and render it another extremely-high-energy \gam-ray source associated to a pulsar \citep[see e.g.][]{Sudoh2021,HAWC2021a,deOnaWilhelmi2022}.
In this context, we also note the recent discovery of extended halos around several energetic pulsars \citep[e.g.][]{HAWC2017}.
While the term `halo' has frequently been adopted in the literature for many \gam-ray sources associated with pulsars \citep[e.g.][]{Linden2017}, we follow here the stricter definition by \citet{Giacinti2020}, who have defined it as a region where the pulsar no longer dominates the dynamics of the interstellar medium (ISM), yet where an over-density of relativistic electrons is present.
The escape of the electrons from the PWN to the extended halo could for example be caused by an interaction of the reverse SNR shock with the pulsar wind \citep{Blondin2001,Hinton2011}.

We present here an updated study of \hessj with \hess, based on a larger data set compared to previous publications, and employing improved analysis methods.
To be able to interpret the results in a consistent manner, we complement this with a new analysis of data above \SI{1}{\GeV} from the \fermilat space telescope for the same region.
In doing so, we are able to gain new insights into the nature of \hessj.

In Sect.~\ref{sec:data_analysis}, we introduce the \hess and \fermilat data sets and analyses.
The results of the analyses are presented in Sect.~\ref{sec:results}, followed by an interpretation in the framework of leptonic and hadronic models in Sect.~\ref{sec:modelling}.
Finally, we conclude the paper in Sect.~\ref{sec:conclusion}.

\section{Data analysis}
\label{sec:data_analysis}

\subsection{\hess data analysis}
\label{sec:hess_analysis}

\hess is an array of five imaging atmospheric Cherenkov telescopes (IACTs), which detect the Cherenkov light produced in atmospheric air showers that are initiated by primary \gam rays.
It is situated on the Southern hemisphere, in the Khomas highlands of Namibia (23\degr{}16\arcmin{}18\arcsec{}S, 16\degr{}30\arcmin{}00\arcsec{}E), at an altitude of \SI{1800}{\meter} above sea level.
The original array, referred to as \mbox{`HESS-I'}, was installed in 2000-2003 and comprised four telescopes with \SI{12}{\meter}-diameter mirrors (CT1-4), arranged in a square layout with \SI{120}{\meter} side length \citep{HESS_Crab_2006}.
The array was completed in 2012 by a fifth telescope, CT5, featuring a \SI{28}{\meter}-diameter mirror, and placed in the centre of the array \citep{Holler2015}.
With the full array, \hess is sensitive to \gam rays in the energy range between $\sim$\SI{0.1}{\TeV} and $\sim$\SI{100}{\TeV}.

\subsubsection{Data set and low-level data analysis}
\label{sec:hess_low_level_analysis}

\hess observations on \hessj have been carried out between 2004 and 2010, that is, exclusively during the HESS-I phase.
The analysis presented here is therefore restricted to the CT1-4 telescopes.
The observations are divided into `runs' of typically \SI{28}{\minute} duration.
Selecting runs that encompass \hessj within $\sim$2.2\degr{} of the pointing position of the telescopes, and applying standard selection criteria for spectral studies \citep{HESS_Crab_2006}, we obtained a data set comprising 201~runs, amounting to a total observation time of \SI{93.2}{\hour}.
This represents a significant increase with respect to the previous dedicated publications on \hessj, which used \SI{25}{\hour} \citep{HESS_J1809_2007}, \SI{32}{\hour} \citep{Komin2007}, and \SI{41}{\hour} \citep{Renaud2008} of data.

In the data analysis, we have selected \gam ray-like events using a machine learning-based method \citep{Ohm2009}, and have reconstructed their energy and arrival direction employing a maximum-likelihood fit, in which the recorded telescope images are compared to a library of simulated image templates \citep{Parsons2014}.
We have repeated the entire analysis with an independent second analysis chain \citep{Becherini2011}, which employs different algorithms for the image calibration, event selection, and event reconstruction, obtaining compatible results.
For the subsequent high-level analysis, we converted our data to the open `GADF' format \citep{Deil2018}, and used the open-source analysis package \gammapy \citep{Deil2017,Deil2020} (v0.17).

Atmospheric air showers initiated by charged cosmic rays outnumber those resulting from \gam rays by several orders of magnitude and they cannot be rejected completely in the event selection without a too severe loss of \gam-ray efficiency.
The modelling of the residual background due to these events (referred to as `hadronic background' hereafter) represents one of the major challenges in any analysis of IACT data.
Most established techniques rely on an estimation of the background from source-free regions in the observed field of view of the run itself \citep[see][for a review]{Berge2007}.
We have chosen here an alternative approach, in which the residual hadronic background is provided by a background model, which we have constructed from archival \hess observations, as detailed in \citet{Mohrmann2019}.
Together with the usage of \gammapy, this enabled us to carry out a 3-dimensional likelihood analysis of the data, that is, to model simultaneously the energy spectrum and spatial morphology of \hessj.
The application of this analysis method to \hess data has been validated by \citet{Mohrmann2019}.

Owing to varying observation conditions (in particular the pointing zenith angle and the atmospheric transparency), a dedicated energy threshold needs to be computed for each observation run.
We determined the thresholds requiring that the average bias of the energy reconstruction does not exceed 10\%, and that the background model is not used below the energy at which the predicted background rate peaks \citep[for more details, see][]{Mohrmann2019}.
The resulting energy thresholds are below \SI{0.9}{\TeV} for all observations, while the lowest threshold, obtained for $\sim$10\% of the observations, is \SI{0.27}{\TeV}.
Because the performance of the system degrades at large offset angles, we furthermore imposed a maximum angle between the direction of reconstructed events and the telescope pointing position of 2.2\degr{}.
The value has been chosen such that the emission region is fully enclosed for all selected observations runs, many of which have been taken as part of the HGPS, implying that a considerable fraction exhibit relatively large (i.e.\ $>1^\circ$) offset angles with respect to the centre of the emission region. 

\subsubsection{Likelihood analysis}
\label{sec:hess_likelihood_analysis}

In the likelihood analysis, the best-fit models have been obtained by minimising the quantity $-2\log(\mathcal{L})$, where $\mathcal{L}=\prod_i P(n_i\,|\,\mu_i)$, and $P(n_i\,|\,\mu_i)$ is the Poisson probability of observing $n_i$ events in bin~$i$, given a predicted number of events $\mu_i$ from the background model, and \gam-ray source models if present \citep{Mattox1996}.
To compute the number of events predicted by source models, we folded the source spatial model and energy spectrum with the instrument response functions (IRFs; effective area, point spread function, and energy dispersion matrix), which we derived for every observation run from extensive Monte Carlo simulations\footnote{Technically, custom IRFs for every observation run are obtained by interpolating between IRFs generated from Monte Carlo simulations that have been carried out for a grid of observational parameters.} \citep{Bernloehr2008}.
As spatial source models we have used 2-dimensional Gaussians that can be either symmetric or elongated, represented by the \texttt{GaussianSpatialModel} class in \gammapy.
As spectral models, we have used a power law (PL) of the form
\begin{linenomath*}
\begin{equation}\label{eq:pl}
  \frac{\mathrm{d}N}{\mathrm{d}E} = N_0\cdot \left(\frac{E}{E_0}\right)^{-\Gamma}\,\,,
\end{equation}
\end{linenomath*}
with normalisation $N_0$, spectral index $\Gamma$, and reference energy $E_0$, as well as a power law with exponential cut-off (ECPL),
\begin{linenomath*}
\begin{equation}\label{eq:ecpl}
  \frac{\mathrm{d}N}{\mathrm{d}E} = N_0\cdot \left(\frac{E}{E_0}\right)^{-\Gamma}\cdot \exp\left(-\frac{E}{E_c}\right)\,\,,
\end{equation}
\end{linenomath*}
where $E_c$ is the cut-off energy.

We carried out the likelihood fit in a region of interest (RoI) of 6\degr{} $\times$ 6\degr{}, centred on \hessj (see Fig.~\ref{fig:sign_map_bkg_fit} in Appendix~\ref{sec:appendix_bkg_fit}).
For the binning of our data, we used spatial pixels of 0.02\degr{} $\times$ 0.02\degr{} size, and an energy binning of 16 bins per decade of energy.
Besides \hessj, the RoI also contains the known \gam-ray sources HESS~J1804$-$216, HESS~J1813$-$178 \citep[][]{HESS_HGPS_2018}, and HESS~J1808$-$204 \citep{HESS_J1808_2018}, which we have masked in the fit using circular exclusion regions (cf.\ Fig.~\ref{fig:sign_map_bkg_fit}).

In the first step of the analysis, we have adjusted the background model for each observation.
This background model fit is described in detail in Appendix~\ref{sec:appendix_bkg_fit}, where we also lay out the procedure for computing significance maps.
The fit result indicates that we have achieved a very good description of the hadronic background after the adjustment.

For the further analysis, we have combined the observations into six `stacked' data sets, where observations with the same energy threshold have been grouped together.
This procedure effectively combines observations with similar observing conditions; further divisions of the data would lead to too many separate data sets.
The six data sets are fitted jointly in the likelihood analysis.
Then, we have modelled the \gam-ray emission of \hessj by adding source components to the model prediction.
For nested\footnote{
  Two models $M_0$ and $M_1$ are considered nested if the parameters of $M_0$ are a subset of those of $M_1$, and $M_1$ can be reduced to $M_0$ for a particular choice of values for its additional parameters.
}
models, the preference of one model over another one can be computed from the `test statistic', $\mathrm{TS}=-2\log(\mathcal{L}_0/\mathcal{L}_1)$, which -- in the limit of sufficient statistics, and far enough from parameter boundaries -- follows a $\chi^2$ distribution with $k$ degrees of freedom, where $k$ is the difference in the number of model parameters between the two tested models \citep{Wilks1938}.

After the model fit, it is possible to extract flux points for each fitted source component.
To do so, we re-ran the fit in narrow energy ranges, keeping all source model parameters except for the flux normalisation $\phi_0$ fixed to their best-fit values.
The best-fit normalisation found in each energy range can then be taken as the measured flux in that range, and is quoted at its centre energy (in logarithmic space).

\subsubsection{Estimation of systematic uncertainties}
\label{sec:hess_sys_error}

Despite the fact that we have computed customised IRFs for each observation run, due to necessary simplifying assumptions in their generation, these IRFs do not always describe the instrument and data-taking conditions perfectly.
Discrepancies between the assumed IRFs and the true conditions can then lead to a systematic bias in the likelihood analysis.
To assess the potential impact of mis-modelled IRFs on our fit results, we have estimated systematic uncertainties for all fit parameters.
Specifically, we have considered two effects that together dominate the systematic uncertainty on our results: a shift of the global energy scale, and uncertainties of the hadronic background model.
A shift of the energy scale may, for example, arise from a mis-modelling of the optical efficiency of the telescopes, or from variations in the transparency for Cherenkov radiation of the atmosphere.
On the other hand, the background model has been constructed from observation runs that were taken under similar, but not identical conditions, and may therefore -- despite its adjustment to the analysed observations (cf.\ Appendix~\ref{sec:appendix_bkg_fit}) -- not predict the background rate perfectly.

\begin{figure*}[th]
  \centering
  \includegraphics{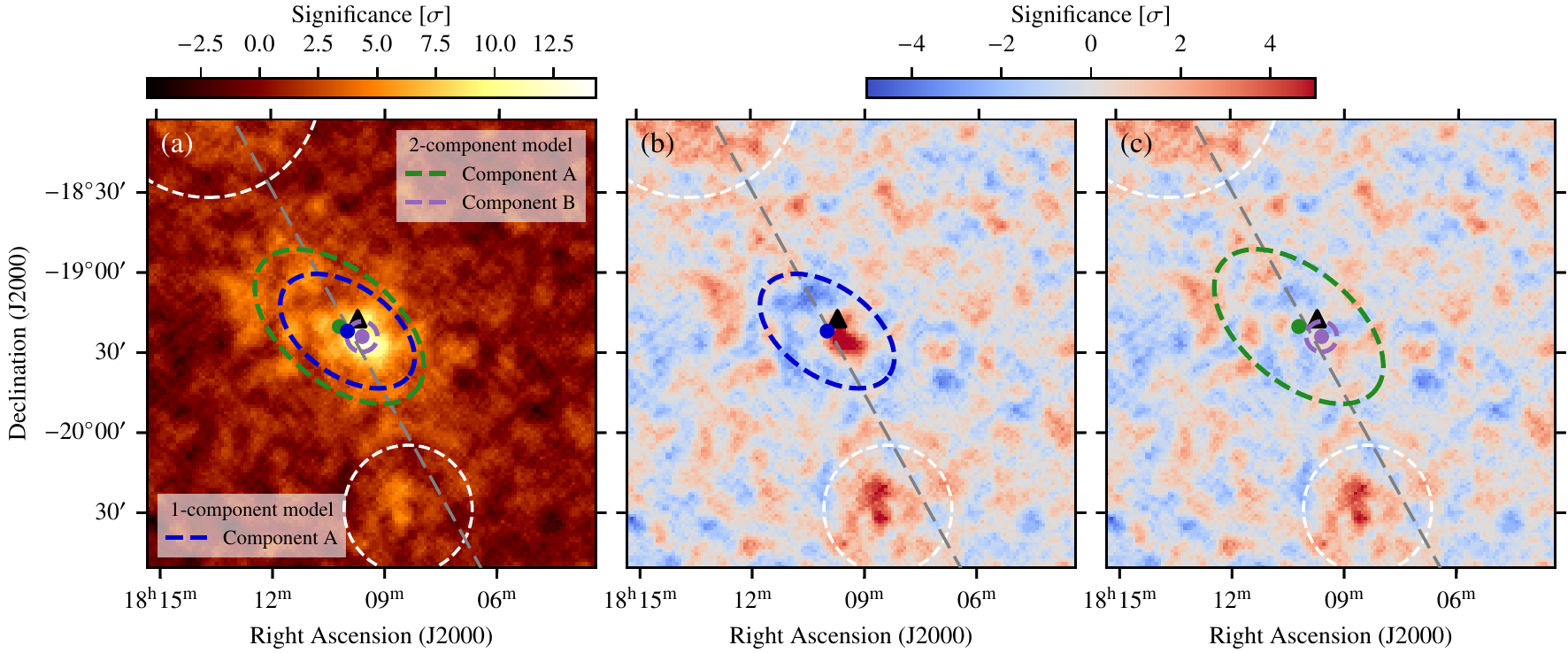}
  \caption{Significance maps with best-fit models.
    We used an oversampling radius of $0.07^\circ$ for smoothing.
    Significance values were obtained using the `Cash' statistic \citep[][see also Appendix~\ref{sec:appendix_bkg_fit}]{Cash1979}.
    Circle markers and coloured dashed lines display the best-fit position and 1-$\sigma$ extent of the Gaussian model components, respectively.
    (a) Pre-modelling significance map, with best-fit model components of the 1-component model and the 2-component model indicated.
    (b) Residual significance map for the 1-component model.
    (c) Residual significance map for the 2-component model.
    In all panels, the grey dashed line marks the Galactic plane, white dashed circles show regions excluded from the analysis, and the black triangle marker denotes the position of \psrj.
  }
  \label{fig:sign_maps_models}
\end{figure*}

We estimated the systematic uncertainties adopting a Monte Carlo-based approach, in which we randomly varied the IRFs according to the two systematic effects mentioned above, generated random pseudo data sets based on these IRFs and the best-fit source models, and re-fitted these pseudo data sets with the original, unmodified IRFs.
The obtained spread in the fitted source model parameters then reflects their combined statistical and systematic uncertainty.
The procedure is described in detail in Appendix~\ref{sec:appendix_sys_err}, and the resulting systematic uncertainties are presented along with the analysis results in Sect.~\ref{sec:hess_results}.

We note that the two considered effects potentially do not encompass all possible sources of systematic errors in the analysis.
For a more general estimate of the systematic uncertainties of \hess, we refer to \citet{HESS_Crab_2006}, where systematic uncertainties of 20\% on the flux normalisation and 0.1 on the source spectral index have been derived.

\subsection{\fermilat data analysis}
\label{sec:fermi_analysis}

\fermilat is a pair conversion telescope onboard the \textit{Fermi} Gamma-Ray Space Telescope and is sensitive to \gam rays in the energy range from $\sim$\SI{20}{\MeV} to several hundred~GeV \citep{FermiLAT2009}.
Here, we analysed 12~years and 5~months of data, taken between August 4, 2008 and December 31, 2020.
We used the `Pass~8' IRFs \citep{Atwood2013} and selected events passing the P8R3\_SOURCE event selection (event class~128, event type~3).
Because the angular resolution of \fermilat substantially worsens below \SI{1}{\GeV}, we restricted the analysis to events above this energy.
To suppress \gam rays originating from the Earth's limb, we furthermore excluded events with zenith angles above $90^\circ$.
The data were analysed using \textsc{Fermitools}\footnote{\url{https://fermi.gsfc.nasa.gov/ssc/data/analysis/software}} version~2.2.0 and \fermipy\footnote{\url{https://fermipy.readthedocs.io}} version~1.1.5 \citep{Wood2017}.

The analysis was carried out in a region of interest (ROI) of $10^\circ\times10^\circ$, centred on the nominal position of 4FGL~\fermijten provided in the 4FGL-DR2 catalogue \citep{FermiLAT_4FGL_2020,FermiLAT_4FGLDR2_2020}.
The events and exposure maps were binned using spatial bins of $0.1^\circ\times0.1^\circ$ size and five bins per decade in energy.
\gam ray source models were then fitted to the data with a likelihood fit as described in Section~\ref{sec:hess_likelihood_analysis}, where we used for the sources in the ROI the models provided in the 4FGL-DR2 catalogue by default.
In addition, we included standard templates for isotropic and Galactic diffuse \gam-ray emission.\footnote{We used the file iso\_P8R3\_SOURCE\_V2\_v1.txt for the isotropic and the file gll\_iem\_v07.fits for the Galactic diffuse emission, respectively; see also \url{ http://fermi.gsfc.nasa.gov/ssc/data/access/lat/BackgroundModels.html}.}

In the likelihood analysis, we have fixed the parameters of all source models with a TS value smaller than~25 or with fewer than~700 predicted events.
We left free the normalisation parameter for all sources within $1^\circ$ of 4FGL~\fermijten, all parameters of the isotropic and Galactic diffuse model, as well as all parameters of the source models for 4FGL~\fermijten and 4FGL~\fermijeleven, which immediately overlap with \hessj.
Systematic uncertainties on the best-fit flux normalisations have been obtained by scaling up and down the effective area by 3\% and repeating the analysis\footnote{We have followed the procedure outlined at \url{https://fermi.gsfc.nasa.gov/ssc/data/analysis/scitools/Aeff_Systematics.html}}.

\section{Results}
\label{sec:results}

\subsection{\hess results}
\label{sec:hess_results}

\begin{figure*}
  \centering
  \includegraphics{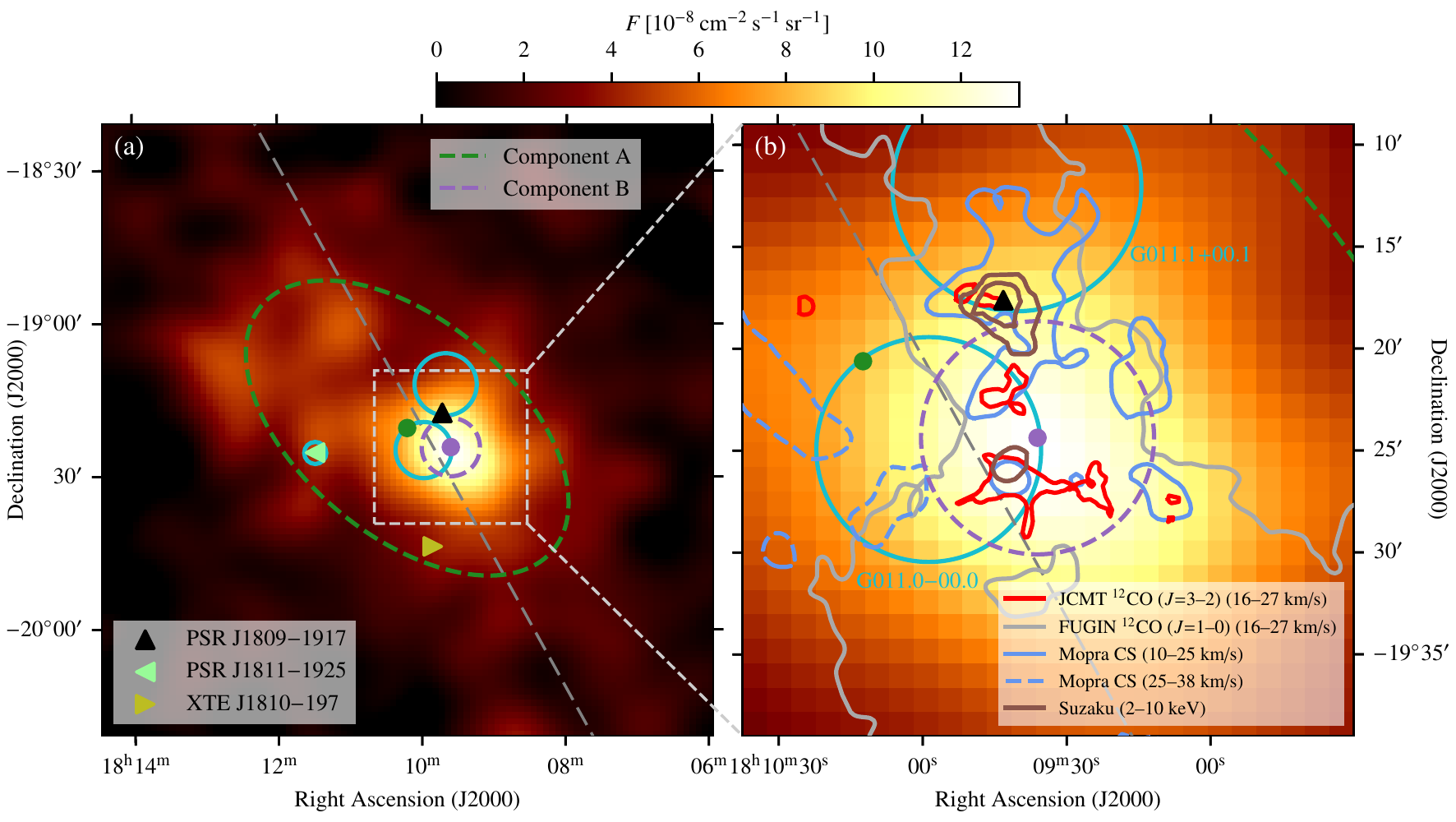}
  \caption{Flux maps for \hessj, above a \gam-ray energy of \SI{0.27}{\TeV}.
    (a) View of the entire emission of \hessj.
    (b) Zoom-in on core region.
    In both panels, light blue circles show the positions of known SNRs, the black triangle marker denotes the position of \psrj, and the grey dashed line marks the Galactic plane.
    The position and extent of component~A and component~B of the two-component model are displayed in green and purple, respectively.
    The multi-wavelength data in panel (b) are from the JCMT \citep{Castelletti2016}, the FUGIN survey \citep{Umemoto2017}, the Mopra telescope \citep{Voisin2019}, and the Suzaku X-ray telescope \citep{Anada2010}.
    The velocity interval for the FUGIN data has been adopted from \citet{Castelletti2016}.
    We computed the flux maps assuming a power law-type spectrum with index $-2.2$, and have employed a Gaussian kernel with $0.07^\circ$ radius for smoothing.
  }
  \label{fig:flux_map}
\end{figure*}

We show in Fig.~\ref{fig:sign_maps_models}(a) the residual significance map after the fit of the hadronic background model, while the map in Fig.~\ref{fig:flux_map}(a) displays the deduced flux of \gam rays above the threshold energy of \SI{0.27}{\TeV}.
Extended \gam-ray emission around the position of \psrj (black triangle marker) is visible, we refer to this source as \hessj.
It is striking that besides the larger-scale emission with an extent of about $1^\circ$, a core of bright emission close to -- but not fully coinciding with -- the pulsar position is present.

\begin{figure}
  \centering
  \includegraphics{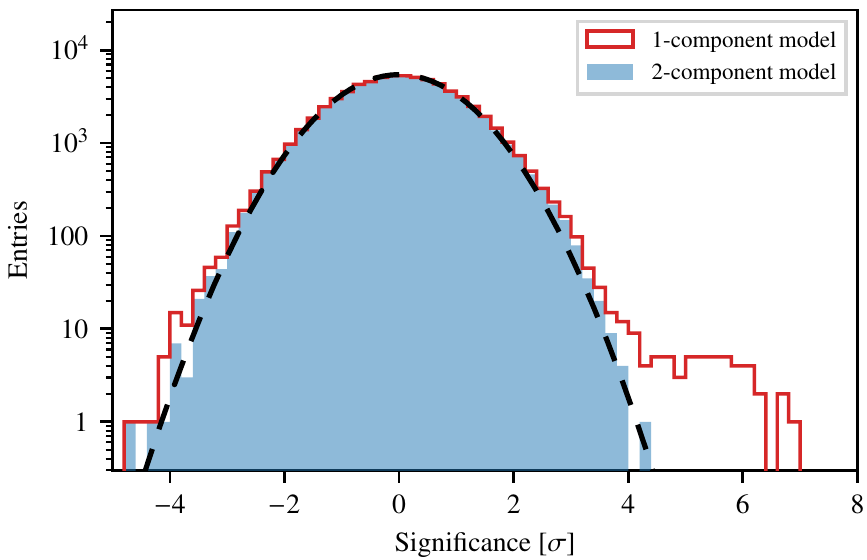}
  \caption{Significance distributions after model fits.
    The red, unfilled histogram shows the distribution for all spatial pixels (outside the exclusion regions) of the significance map in Fig.~\ref{fig:sign_maps_models}(b), while the blue, filled histogram corresponds to the map in Fig.~\ref{fig:sign_maps_models}(c).
    The black, dashed line displays a Gaussian distribution with a mean $\mu=0$ and width $\sigma=1$.
  }
  \label{fig:sign_dist_models}
\end{figure}

First, we have attempted to model the emission using a 1-component model, which comprises a single source component, described by an elongated Gaussian spatial model and a PL spectral model.
The best-fit position and 1-$\sigma$ extent of this component are shown in blue in Fig.~\ref{fig:sign_maps_models}(a), and in Fig.~\ref{fig:sign_maps_models}(b) we show the residual significance map after subtracting the emission predicted by this model.
The residual map shows significant remaining features close to the best-fit position, indicating that the 1-component model does not provide an acceptable description of the observed emission.
This finding is confirmed by the corresponding distribution of pixel significance values, shown by the red line in Fig.~\ref{fig:sign_dist_models}, which clearly deviates from the expected distribution for a good description of the data (black, dashed line).
This is because the larger-scale emission and the bright core cannot be simultaneously modelled by a single component that is described by a Gaussian, or any other reasonably basic spatial model.
This finding holds even if we allow the extent of the Gaussian model to vary with energy, as detailed in Appendix~\ref{sec:appendix_1comp_fit_ebands}.

We have therefore adopted a 2-component model, which features in addition a second component that is described by a symmetric Gaussian spatial model and a PL spectral model.
Fig.~\ref{fig:sign_maps_models}(a) also shows the best-fit position and extent of both components of the 2-component model, while the residual significance map after the fit of this model is displayed in Fig.~\ref{fig:sign_maps_models}(c), and the corresponding distribution of significance values is shown in Fig.~\ref{fig:sign_dist_models} (blue histogram).
As is clearly visible, the 2-component model provides a much better fit to the observed data than the 1-component model (statistically, it is preferred by $13.3 \sigma$), and in fact no residual deviations except for those expected from statistical fluctuations can be made out.
In the following, we will refer to the extended component as `component~A', and to the compact component as `component~B'.
In Appendix~\ref{sec:appendix_2comp_maps_ebands}, we study the agreement between the 2-component model and the observed data as a function of energy, finding that the model provides a good fit at all energies.
Finally, we have explored in Appendix~\ref{sec:appendix_2comp_fit_ebands} how the parameters of component~A vary when the model is fitted in separate energy bands.
We find that the fitted parameters do not change significantly, and provide in Table~\ref{tab:size_comp1} the fitted extent of component~A in the four employed energy bands.

\begin{table}
  \centering
  \caption{Extent of \hess component~A in energy bands.}
  \label{tab:size_comp1}
  \begin{tabular}{cccc}
    \hline\hline
    $E_\mathrm{min}-E_\mathrm{max}$ & $E_\mathrm{mean}$ & $\sigma_\mathrm{major}$ & $\sigma_\mathrm{minor}$\\
    (TeV) & (TeV) & (deg) & (deg)\\
    \hline
    $0.27 - 0.75$ & 0.43 & $0.69\pm 0.10$ & $0.41 \pm 0.10$ \\
    $0.75 - 2.1$ & 1.2 & $0.62\pm 0.05$ & $0.31 \pm 0.04$ \\
    $2.1 - 5.6$ & 3.2 & $0.62\pm 0.06$ & $0.35 \pm 0.05$ \\
    $>5.6$ & 9.6 & $0.61\pm 0.11$ & $0.46 \pm 0.13$ \\
    \hline
  \end{tabular}
  \tablefoot{
    $E_\mathrm{min}$ and $E_\mathrm{max}$ denote the lower and upper boundary of the energy band, respectively, $E_\mathrm{mean}$ the weighted mean energy.
    $\sigma_\mathrm{major}$ and $\sigma_\mathrm{minor}$ denote the 1-sigma extent of the semi-major and semi-minor axis of the elongated Gaussian spatial model for \hess component~A, respectively.
    See Appendix~\ref{sec:appendix_2comp_fit_ebands} for further details.
  }
\end{table}

In Fig.~\ref{fig:flux_map}(b), we provide a detail view of the inner region of \hessj, with multi-wavelength data overlaid.
The peak of the emission (and, by that, the position of component~B), is offset by $\sim$7\arcmin{} from the position of \psrj and its surrounding X-ray nebula, indicated by the brown contour lines \citep{Anada2010}.
On the other hand, component~B lies with its centre point directly on the western edge of the SNR~G011.0+00.0, and is furthermore spatially coincident with dense molecular clouds as indicated by $^{12}$CO ($J$=3--2) observations by the James Clerk Maxwell Telescope \citep[JCMT;][]{Castelletti2016} and CS observations by the Mopra telescope \citep{Voisin2019}.
Moreover, the contour lines from the FUGIN~$^{12}$CO ($J$=1--0) survey \citep{Umemoto2017} illustrate that molecular gas is present throughout the region.
In Appendix~\ref{sec:appendix_fugin_map}, we provide a map of the FUGIN $^{12}$CO data with the two components of \hessj overlaid.

\begin{figure}
  \centering
  \includegraphics{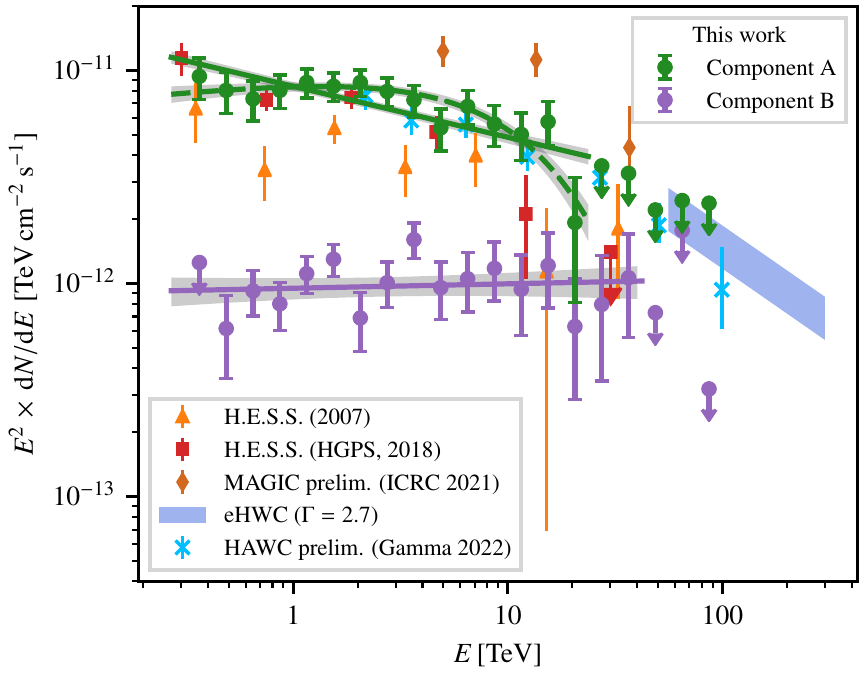}
  \caption{\hess energy spectrum results.
    We show in green and purple the flux points for component~A and component~B of the \hess analysis, respectively.
    Upper limits are at 95\% confidence level.
    The solid lines with shaded bands display the best-fit PL model and statistical uncertainty for each of the components.
    For component~A, the dashed green line shows the best-fit ECPL model in addition.
    The energy spectra are compared to published results, taken from \citet{HESS_J1809_2007}, \citet{HESS_HGPS_2018}, \citet{Zaric2021}, \citet{HAWC2020}, and \citet{Goodman2022}.
  }
  \label{fig:sed_hess}
\end{figure}

Finally, we show in Fig.~\ref{fig:sed_hess} the energy spectra and flux points obtained for the two components of \hessj, and compare these to previously obtained results from the literature.
We note that, when comparing the fitted PL models for component~A and~B (solid lines), the spectrum of component~B appears somewhat harder than that of component~A ($\Gamma=1.98\pm 0.05$ vs. $\Gamma=2.24\pm 0.03$).
However, the flux upper limits obtained above energies of $\sim$\SI{20}{\TeV} for component~A seem to indicate the presence of a cut-off to the spectrum.
We have therefore repeated the analysis adopting an ECPL spectral model for component~A, which led to no changes in the best-fit parameters values for component~B or for the spatial model of component~A, but yielded a flatter spectrum ($\Gamma=1.90\pm 0.05$) also for component~A at energies below $\sim$\SI{5}{\TeV} (dashed line in Fig.~\ref{fig:sed_hess}).
With respect to the PL spectral model, the ECPL model is preferred with a statistical significance of $\sim$8$\sigma$.
For component~B, on the other hand, we found no significant preference for a cut-off to the spectrum.

The spectrum of component~A, which dominates the total emission except at the highest energies, is well compatible with that published in \citet{HESS_HGPS_2018}, and may also be reconcilable with the high-energy emission measured with HAWC \citep{HAWC2020,Goodman2022}, although a more gradual decrease of the flux than predicted by the ECPL spectral model would be required in this case.
That the flux points from \citet{HESS_J1809_2007} indicate a lower flux compared to the one found here can be understood when considering that the flux was extracted only from within a circular region of $0.5^\circ$ radius, and thus part of the larger-scale emission has been missed.

We summarise the best-fit parameter values found for the 2-component model in Table~\ref{tab:model_pars}, along with the corresponding statistical and systematic uncertainties (the latter having been derived as described in Appendix~\ref{sec:appendix_sys_err}).
For component~A, we provide the results both for the PL spectral model and the ECPL spectral model.

\begin{table}
  \centering
  \caption{Best-fit parameter values for the \hess 2-component model.
    For component~A, we provide the best-fit values for the assumption of a PL spectral model and an ECPL spectral model. 
  }
  \label{tab:model_pars}
  \begin{tabular}{lc}
    \hline\hline
    Par. [unit] & Value \\\hline
    \multicolumn{2}{c}{\rule{0pt}{1.1\normalbaselineskip} Component~A (PL spectral model) \vspace{0.05cm}}\\\hline
    R.A. [deg] & $272.551\pm 0.025_\mathrm{stat}\pm 0.018_\mathrm{sys}$\\
     & ($18^\mathrm{h}10^\mathrm{m}12^\mathrm{s} \pm 6^\mathrm{s}_\mathrm{stat} \pm 4^\mathrm{s}_\mathrm{sys}$)\\
    Dec. [deg] & $-19.344\pm 0.023_\mathrm{stat}\pm 0.013_\mathrm{sys}$\\
     & ($-19^\circ 20.6'\pm 1.4'_\mathrm{stat}\pm 0.8'_\mathrm{sys})$\\
    $\sigma$ [deg] & $0.622\pm 0.032_\mathrm{stat}\pm 0.020_\mathrm{sys}$\tablefootmark{a}\\
    $e$ & $0.824\pm 0.025_\mathrm{stat}$\\
    $\phi$ [deg] & $50.0\pm 3.1_\mathrm{stat}$\\
    $N_0\,[10^{-12}\,\mathrm{TeV}^{-1}\,\mathrm{cm}^{-2}\,\mathrm{s}^{-1}]$ & $8.42\pm 0.40_\mathrm{stat}\pm 1.14_\mathrm{sys}$\\
    $\Gamma$ & $2.239\pm 0.027_\mathrm{stat}\pm 0.020_\mathrm{sys}$\\
    $E_0$ [TeV] & 1 (fixed)\\\hline
    \multicolumn{2}{c}{\rule{0pt}{1.1\normalbaselineskip} Component~A (ECPL spectral model) \vspace{0.05cm}}\\\hline
    R.A. [deg] & $272.554\pm 0.025_\mathrm{stat}\pm 0.019_\mathrm{sys}$\\
     & ($18^\mathrm{h}10^\mathrm{m}13^\mathrm{s} \pm 6^\mathrm{s}_\mathrm{stat} \pm 5^\mathrm{s}_\mathrm{sys}$)\\
    Dec. [deg] & $-19.344\pm 0.021_\mathrm{stat}\pm 0.012_\mathrm{sys}$\\
     & ($-19^\circ 20.6'\pm 1.3'_\mathrm{stat}\pm 0.7'_\mathrm{sys})$\\
    $\sigma$ [deg] & $0.613\pm 0.031_\mathrm{stat}\pm 0.015_\mathrm{sys}$\tablefootmark{a}\\
    $e$ & $0.820\pm 0.025_\mathrm{stat}$\\
    $\phi$ [deg] & $51.3\pm 3.1_\mathrm{stat}$\\
    $N_0\,[10^{-12}\,\mathrm{TeV}^{-1}\,\mathrm{cm}^{-2}\,\mathrm{s}^{-1}]$ & $9.05\pm 0.47_\mathrm{stat}\pm 0.91_\mathrm{sys}$\\
    $\Gamma$ & $1.90\pm 0.05_\mathrm{stat}\pm 0.05_\mathrm{sys}$\\
    $E_c$ [TeV] & $\left.12.7_{-2.1}^{+2.7}\right|_\mathrm{stat}\left._{-1.9}^{+2.6}\right|_\mathrm{sys}$\\
    $E_0$ [TeV] & 1 (fixed)\\\hline
    \multicolumn{2}{c}{\rule{0pt}{1.1\normalbaselineskip} Component~B \vspace{0.05cm}}\\\hline
    R.A. [deg] & $272.400\pm 0.010_\mathrm{stat}$\\
     & ($18^\mathrm{h}09^\mathrm{m}36^\mathrm{s} \pm 2.4^\mathrm{s}_\mathrm{stat}$)\\
    Dec. [deg] & $-19.406\pm 0.009_\mathrm{stat}$\\
     & ($-19^\circ 24.4'\pm 0.5'_\mathrm{stat})$\\
    $\sigma$ [deg] & $0.0953\pm 0.0072_\mathrm{stat}\pm 0.0034_\mathrm{sys}$\\
    $N_0\,[10^{-12}\,\mathrm{TeV}^{-1}\,\mathrm{cm}^{-2}\,\mathrm{s}^{-1}]$ & $0.95\pm 0.11_\mathrm{stat}\pm 0.011_\mathrm{sys}$\\
    $\Gamma$ & $1.98\pm 0.05_\mathrm{stat}\pm 0.03_\mathrm{sys}$\\
    $E_0$ [TeV] & 1 (fixed)\\\hline
  \end{tabular}
  \tablefoot{
    $\sigma$, $e$, and $\phi$ denote the 1-$\sigma$ radius, eccentricity, and position angle of the Gaussian spatial model, respectively.
    $N_0$, $\Gamma$, $E_0$, and $E_c$ are parameters of the spectral models, as defined in Eqs.~(\ref{eq:pl}) and (\ref{eq:ecpl}).
    Systematic uncertainties have been derived as described in Appendix~\ref{sec:appendix_sys_err}.\\
    \tablefoottext{a}{For the asymmetric component~A, $\sigma$ refers to the semi-major axis. The extents of the semi-minor axis compute to $0.353 \pm 0.029_\mathrm{stat}$ and $0.351 \pm 0.028_\mathrm{stat}$ for the PL and ECPL spectral model, respectively.}
  }
\end{table}

\subsection{\fermilat results}
\label{sec:fermi_results}

The \fermilat 4FGL-DR2 catalogue \citep{FermiLAT_4FGL_2020,FermiLAT_4FGLDR2_2020} lists two sources that are located in the immediate vicinity of \hessj:\footnote{We have used the 4FGL-DR2 catalogue as a basis for our analysis, but have checked that the region is modelled in the same way in the more recent 4FGL-DR3 catalogue \citep{FermiLAT_4FGLDR3_2022}.}
(i) 4FGL~\fermijten is modelled as an extended source (using a two-dimensional Gaussian as spatial model) and its spectrum is fitted with a log-parabola model,
\begin{linenomath*}
\begin{equation}\label{eq:logparabola}
  \frac{\mathrm{d}N}{\mathrm{d}E} = N_0\cdot \left(\frac{E}{E_0}\right)^{-\alpha-\beta\log(E/E_0)}\,\,
\end{equation}
\end{linenomath*}
(with log the natural logarithm);
(ii) 4FGL~\fermijeleven is modelled as a point-like source and its spectrum is fitted with a power-law model (cf.\ Eq.~\ref{eq:pl}).
We confirm with our analysis the presence of both sources, that is, we were not able to obtain a satisfactory fit with only one source, or a different choice of spatial models.
In the following, we will refer to the source models we obtained as \fermijten and \fermijeleven, respectively, in distinction to the models provided in the 4FGL-DR2 catalogue.
The best-fit parameter values of the models are summarised in Table~\ref{tab:fermi_model_pars}.
The systematic uncertainties on all model parameters except the flux normalisation $N_0$ are negligible compared to the statistical ones, and thus not quoted in the table.
We acknowledge that our best-fit spectral model for \fermijten shows no significant curvature, but have decided to maintain the log-parabola model for consistency with the 4FGL-DR2 catalogue (which extended to lower energies than our analysis here).

\begin{table}
  \centering
  \caption{Best-fit parameter values for the \fermilat data analysis.
  }
  \label{tab:fermi_model_pars}
  \begin{tabular}{lc}
    \hline\hline
    Par. [unit] & Value \\\hline
    \multicolumn{2}{c}{\rule{0pt}{1.1\normalbaselineskip} \fermijten \vspace{0.05cm}}\\\hline
    R.A. [deg] & $272.547\pm 0.033_\mathrm{stat}$\\
     & ($18^\mathrm{h}10^\mathrm{m}11^\mathrm{s} \pm 8^\mathrm{s}_\mathrm{stat}$)\\
    Dec. [deg] & $-19.397\pm 0.038_\mathrm{stat}$\\
     & ($-19^\circ 23.8'\pm 2.3'_\mathrm{stat})$\\
    $\sigma$ [deg] & $0.317\pm 0.024_\mathrm{stat}$\\
    $N_0\,[10^{-9}\,\mathrm{GeV}^{-1}\,\mathrm{cm}^{-2}\,\mathrm{s}^{-1}]$ & $2.86\pm 0.27_\mathrm{stat}\pm 0.09_\mathrm{sys}$\\
    $\alpha$ & $2.53\pm 0.11_\mathrm{stat}$\\
    $\beta$ & $-0.015\pm 0.043_\mathrm{stat}$\\
    $E_0$ [GeV] & 1.747 (fixed)\\\hline
    \multicolumn{2}{c}{\rule{0pt}{1.1\normalbaselineskip} \fermijeleven \vspace{0.05cm}}\\\hline
    R.A. [deg] & $272.874\pm 0.025_\mathrm{stat}$\\
     & ($18^\mathrm{h}11^\mathrm{m}30^\mathrm{s} \pm 6^\mathrm{s}_\mathrm{stat}$)\\
    Dec. [deg] & $-19.410\pm 0.027_\mathrm{stat}$\\
     & ($-19^\circ 24.6'\pm 1.6'_\mathrm{stat})$\\
    $N_0\,[10^{-9}\,\mathrm{GeV}^{-1}\,\mathrm{cm}^{-2}\,\mathrm{s}^{-1}]$ & $0.016\pm 0.003_\mathrm{stat}\pm 0.001_\mathrm{sys}$\\
    $\Gamma$ & $2.42\pm 0.16_\mathrm{stat}$\\
    $E_0$ [GeV] & 7.48 (fixed)\\\hline
  \end{tabular}
  \tablefoot{
    $\sigma$ is the 1-$\sigma$ radius of the Gaussian spatial model.
    $N_0$, $\alpha$, $\beta$, $\Gamma$, and $E_0$ are parameters of the spectral models defined in Eqs.~(\ref{eq:logparabola}) and (\ref{eq:pl}).
  }
\end{table}

\begin{figure*}
  \centering
  \includegraphics{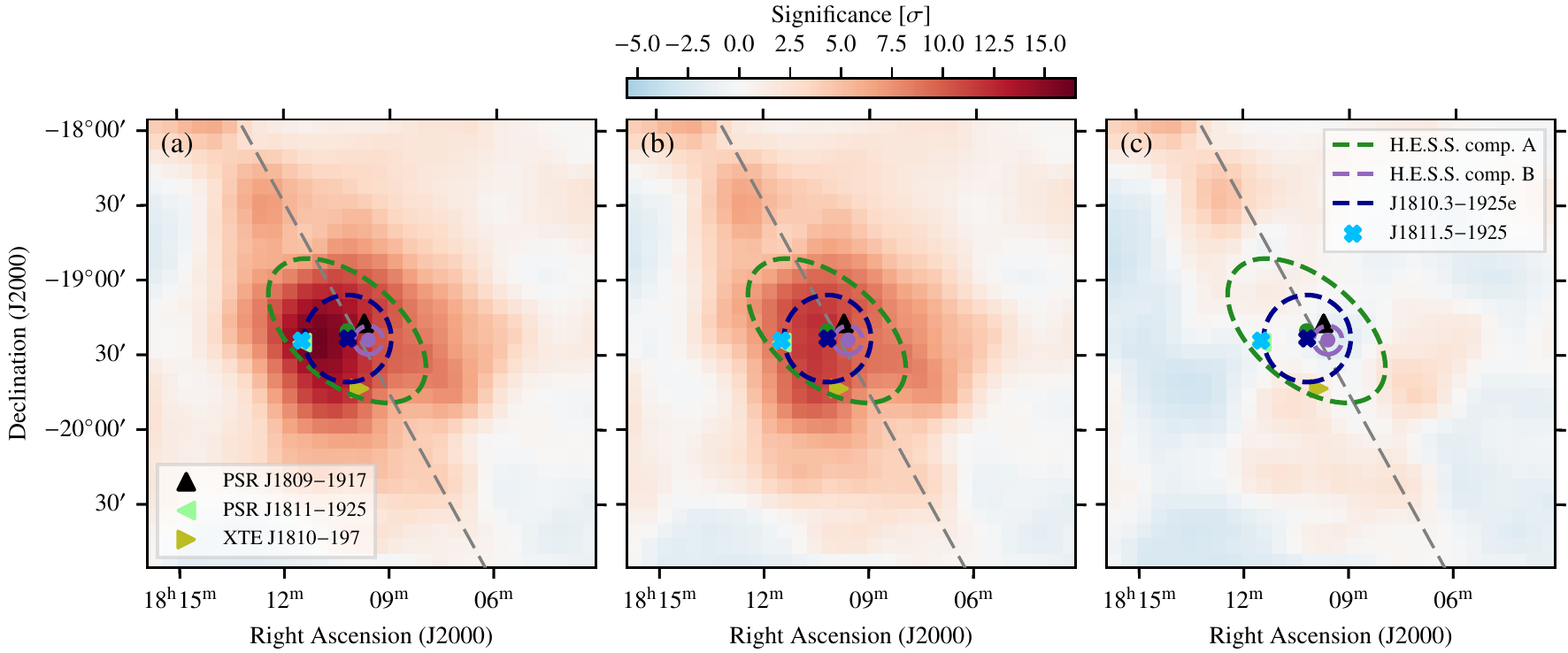}
  \caption{
    Significance maps above \SI{1}{\GeV} from the \fermilat analysis.
    (a) With \fermijeleven and \fermijten removed from the best-fit model.
    (b) With \fermijten removed from the best-fit model.
    (c) With all sources included in the model.
    The light blue cross denotes the fitted position of \fermijeleven, whereas the dark blue cross and dashed circle display the fitted position and 1-$\sigma$ extent of \fermijten.
    For comparison, the components of the 2-component model fitted to the \hess data are shown in green and purple (same as in Fig.~\ref{fig:sign_maps_models}).
    The grey dashed line marks the Galactic plane, while the coloured triangle markers denote the positions of \psrj, \psrjother, and XTE~J1810$-$197.
  }
  \label{fig:fermi_sign_maps}
\end{figure*}

\begin{figure}
  \centering
  \includegraphics{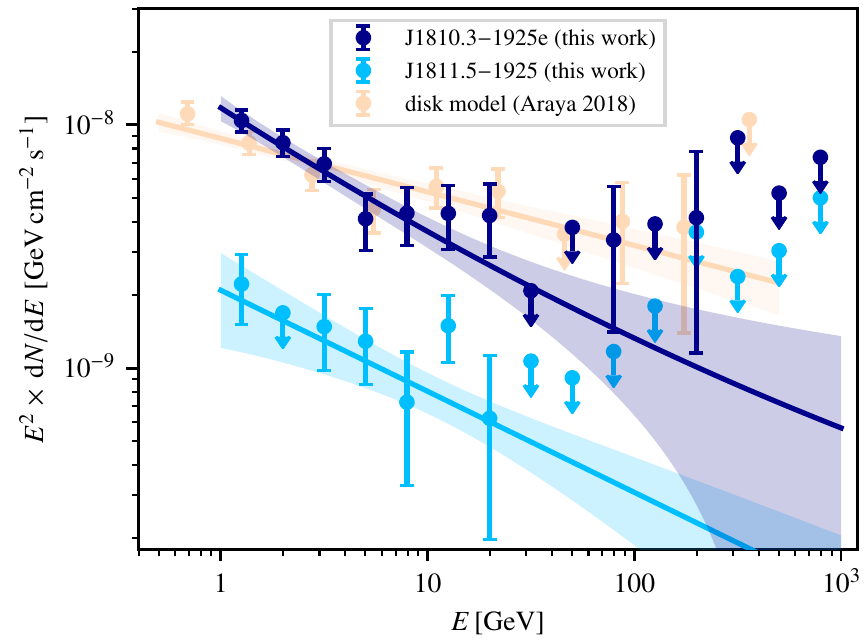}
  \caption{
    \fermilat energy spectrum results for \fermijten and \fermijeleven.
    We compare our spectra to that obtained by \citet{Araya2018} for the entire region.
  }
  \label{fig:sed_fermi}
\end{figure}

Removing \fermijten and \fermijeleven from the best-fit ROI model, we obtained the significance map shown in Fig.~\ref{fig:fermi_sign_maps}, panel~(a).
Panel~(b) displays the significance map after adding \fermijeleven to the ROI model, while panel~(c) shows the map with both source models restored.
A comparison of the panels shows that the two fitted sources account for the majority of the emission around \hessj.
The energy spectra extracted for \fermijten and \fermijeleven are displayed in Fig.~\ref{fig:sed_fermi}.
The obtained spectrum for \fermijten is in good agreement with that obtained by \citet{Araya2018}, considering that the region has been modelled differently (a single source with a disk spatial model) there.

\begin{figure}
  \centering
  \includegraphics{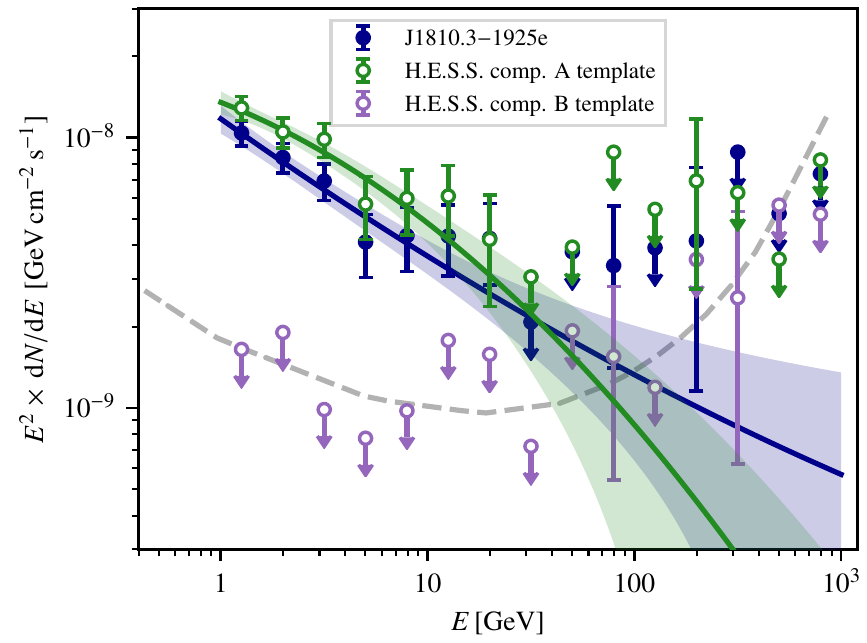}
  \caption{
    \fermilat energy spectra obtained with \hess model templates.
    The dashed grey line shows the 10-year \fermilat broadband sensitivity.
  }
  \label{fig:sed_fermi_hess_comp}
\end{figure}

The point-like source \fermijeleven is positionally coincident with \psrjother, which strongly suggests an association with this pulsar, as also listed in the 4FGL-DR2 catalogue.
We therefore regard its emission as unrelated to \hessj.
On the other hand, the best-fit position of \fermijten is close to \psrj and the two \hess source components, suggesting a connection to \hessj.
In particular, the fitted position and extent are very similar to those of the extended \hess component (component~A), as is evident from Fig.~\ref{fig:fermi_sign_maps}.
In order to further explore the connection between the \fermilat and \hess data, we have also extracted energy spectra of the emission observed with \fermilat using the best-fit spatial models of the two \hess components as spatial templates (removing \fermijten from the model but retaining \fermijeleven).
The result is shown in Fig.~\ref{fig:sed_fermi_hess_comp}.
As expected due to its slightly larger spatial extent, the spectrum obtained for the template of component~A is slightly above that of \fermijten.
With the template of component~B we obtained only flux upper limits, however, this is not a surprise given \fermilat's broadband sensitivity\footnote{Broadband sensitivity curves for \fermilat are available at \url{https://www.slac.stanford.edu/exp/glast/groups/canda/lat_Performance}. We have used the curve for Galactic coordinates $l=0^\circ$ and $b=0^\circ$.} (dashed line in Fig.~\ref{fig:sed_fermi_hess_comp}).

\section{Modelling}
\label{sec:modelling}

In this section, we present an interpretation of the observational results by means of modelling the primary cosmic-ray particle populations responsible for the observed \gam-ray emission.
We investigate two scenarios: (i) that the emission detected with \hess is entirely attributed to the PWN of \psrj, that is, of purely leptonic origin (Section~\ref{sec:pwn_scenario}); (ii) that there is an additional contribution to the emission from hadronic cosmic rays accelerated in one of the SNRs and interacting in the molecular clouds (Section~\ref{sec:hadronic_scenario}).

\subsection{Pulsar wind nebula scenario}
\label{sec:pwn_scenario}

We employed a one-zone PWN model, in which we performed a time-dependent modelling of the pulsar energy output, the ambient magnetic field, and the injected electrons, following the approach outlined in \citet{HAWC2021}.
The parameters of the model are summarised in Table~\ref{tab:pars_pwn_model}.
The input parameters consist mostly of measured properties of \psrj.
For the pulsar braking index, however, we have adopted the canonical value of $n=3$, assuming that this is more representative for the full history of the pulsar than the recent measurement of $n=23.5$, which may be affected by undetected glitches of the pulsar \citep{Parthasarathy2019,Parthasarathy2020}.
The electron injection spectrum follows a power law with index $-\alpha$ and an exponential cut-off at energy $E_c$.
Its normalisation is proportional to $\theta\times \dot{E}$, that is, coupled to the time-dependent spin-down power.
As specified in \citet{Gaensler2006} and \citet{Venter2007}, we took the time evolution of the pulsar period as $P(t)=P_0(1+t/\tau_0)^{0.5}$, of the pulsar spin-down power as $\dot{E}(t)=\dot{E}_0(1+t/\tau_0)^{-2}$, and of the magnetic field as $B(t)=B_0[1+(t/\tau_0)^{0.5}]^{-1}$, where $\tau_0=P_0/(2\dot{P}_0)$ is the initial spin-down time scale.
We then computed the non-thermal emission from the injected electrons (i.e.\ synchrotron radiation and IC emission) employing the GAMERA library \citep{Hahn2015}, which takes into account cooling losses of the electrons.
For the IC target photon fields, we have used the model by \citet{Popescu2017}\footnote{We note that the prediction of this large-scale model may not be very accurate in the specific region studied here. As our conclusions are based on order-of-magnitude estimates, however, they are unaltered even if the predicted radiation field densities are wrong by a factor of a few.}.
Finally, we have fitted the adjustable parameters of the model to the observed spectral energy distribution (SED) of \hessj, where the optimisation has been carried out using a Markov chain Monte Carlo (MCMC) method implemented in the \textsc{emcee} package \citep{ForemanMackey2013}.
We note that some of the model parameters are correlated with each other or not well constrained by the available data.
Therefore, we stress that while we have carried out an optimisation of the model parameters, the obtained values should not be regarded as measurements of the corresponding quantities, but rather as one possible combination of parameter values that yield a reasonable description of the observational data.
The parameter values given in Table~\ref{tab:pars_pwn_model} are those that yielded the highest numerical probability, that is, the best fit to the data.

\begin{table}
  \centering
  \caption{Parameters of the PWN model}
  \label{tab:pars_pwn_model}
  \begin{tabular}{ccc}
    \hline\hline
    Par. & Description & Value\\
    \hline
    \multicolumn{3}{c}{\rule{0pt}{1.1\normalbaselineskip} Input parameters \vspace{0.05cm}}\\\hline
    $d$ & pulsar distance\tablefootmark{a} & \SI{3.3}{\kpc}\\
    $\dot{E}$ & pulsar spin-down power\tablefootmark{a} & \SI{1.8e36}{\erg\per\second}\\
    $\tau_c$ & pulsar characteristic age\tablefootmark{a} & \SI{51.4}{\kilo\year}\\
    $P$ & pulsar period \tablefootmark{a} & \SI{82.76}{\milli\second}\\
    $\dot{P}$ & pulsar period derivative\tablefootmark{a} & \SI{2.55e-14}{\second\per\second}\\
    $n$ & pulsar braking index\tablefootmark{b} & 3\\
    \hline
    \multicolumn{3}{c}{\rule{0pt}{1.1\normalbaselineskip} Adjusted parameters \vspace{0.05cm}}\\\hline
    $\theta$ & electron power fraction & $0.6$\\
    $B$ & magnetic field & $\SI{4}{\micro\gauss}$\\
    $P_0$ & pulsar birth period & $\SI{50}{\milli\second}$\\
    $E_c$ & cut-off energy & $\SI{420}{\TeV}$\\
    $\alpha$ & injection spectrum index & $2.0$\\
    $\tau_\mathrm{young}$ & age of young e$^-$ & $\SI{1.2}{\kilo\year}$\\
    $\tau_\mathrm{med}$ & age of medium-age e$^-$ & $\SI{4.7}{\kilo\year}$\\
    \hline
    \end{tabular}
    \tablefoot{
      Pulsar parameters denote present-day values unless otherwise specified.
      The parameter values for the `adjusted' parameters were obtained using an MCMC method, but should be regarded as indicative values rather than precise fit results (see main text).\\
      \tablefoottext{a}{Taken from \citet{Manchester2005}.}
      \tablefoottext{b}{Assumed value.}
    }
\end{table}

In the model, we invoked three `generations' of electrons:
(i)~`relic' electrons, which have been injected over the life time of the system ($\tau\approx \SI{33}{\kilo\year}$\footnote{The `true' age of the pulsar can be computed as $\tau=(P/((n-1)\cdot \dot{P}))\cdot(1-(P_0/P)^{n-1})$ \citep{Gaensler2006}.
We note that this formula depends on the unknown pulsar birth period $P_0$, for which we have used the value suggested by our model optimisation (cf.\ Table~\ref{tab:pars_pwn_model}).
Other values of $P_0$ will lead to different estimates of the pulsar age.
})
and are associated with the extended \hess component (A);
(ii)~`medium-age' electrons, which have been injected within the last $\tau_\mathrm{med}\approx \SI{4.7}{\kilo\year}$ and are associated with the compact \hess component (B);
(iii)~`young' electrons, which have been injected within the last $\tau_\mathrm{young}\approx \SI{1.2}{\kilo\year}$ and are associated with the X-ray nebula.
In this picture, the `relic' electrons are assumed to have escaped from the central region (which contains the X-ray PWN and the compact component~B) at some instant in the past.
For lack of evidence when this escape has occurred, the `relic' electrons are injected from the birth of the system until the `medium-age' electrons start to be injected.
We note that, despite associating the different generations with different spatial regions, we have not performed a spatial modelling -- the association is made in terms of the SED only.

In addition to the already presented \hess spectra, we used in the fit the spectrum of the X-ray nebula as measured by \citet{Anada2010} with the Suzaku satellite between 2~and \SI{10}{\keV}.
Since we associate only the most recently injected `young' electrons with the X-ray nebula, we integrated the measured flux in the immediate vicinity of the nebula only (regions 2, 3, 6, and 7 in Fig.~4 / Table~4 of \citeauthor{Anada2010} \citeyear{Anada2010}).
Additionally, we derived an upper limit (at 95\% confidence level) for the X-ray flux emitted by the `medium-age' electrons using the measured flux in regions 9--16 and applying a scaling factor that takes into account the difference in solid angle between these regions and the compact \hess component associated with the `medium-age' electrons.
The upper limit is not used in the fit and only serves as a sanity check for the model.

\begin{figure*}
  \centering
  \subfigure[]{
    \includegraphics{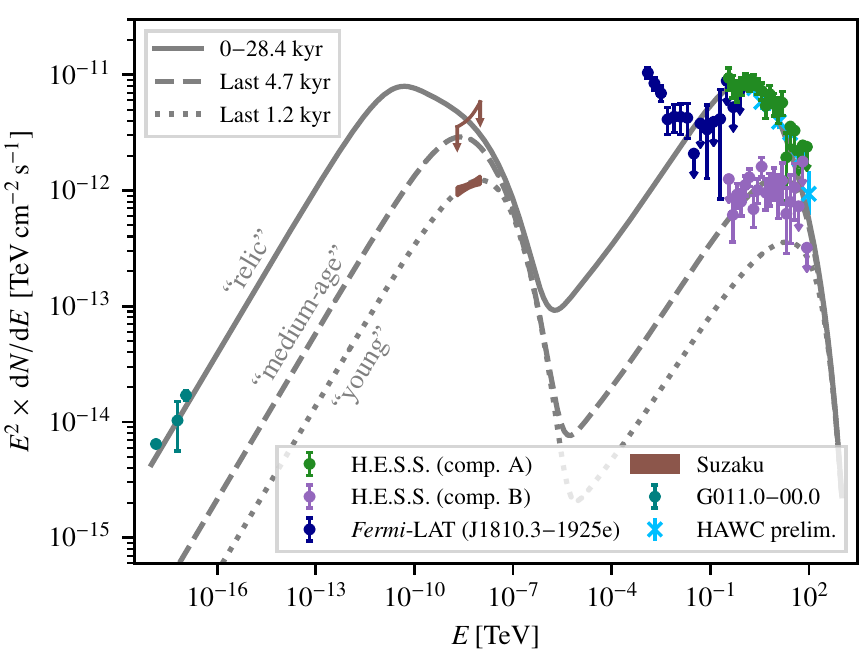}
    \label{fig:model_pwn_full}
  }
  \subfigure[]{
    \includegraphics{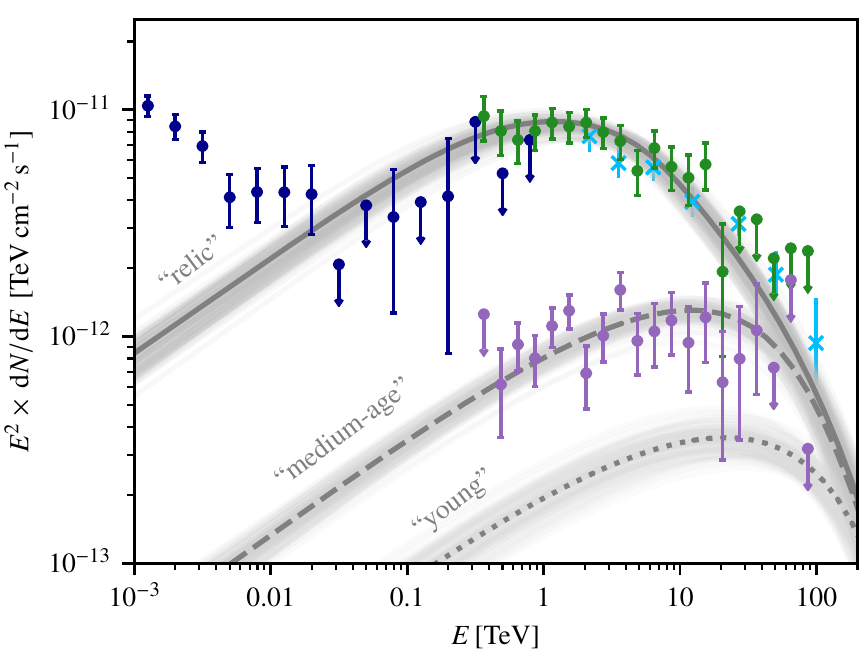}
    \label{fig:model_pwn_zoom}
  }
  \caption{
    SED of \hessj for the leptonic model.
    (a) Full energy range.
    (b) Zoom into high-energy regime.
    SED curves for the three assumed electron generations, obtained with GAMERA, are shown with dark grey lines.
    In panel (b), the light grey lines show individual solutions from the MCMC sampling, and thus give an indication of the statistical spread.
    The \hess and \fermilat data points have been derived in this work.
    The Suzaku X-ray data are from \citet{Anada2010}, where the butterfly corresponds to the `young' electrons (dotted line) and the upper limit refers to the `medium-age' electrons (dashed line; see main text for details).
    Shown for comparison but not used in the fit are radio data for \snrS \citep{Brogan2006} and data points from HAWC \citep{Goodman2022}.
  }
  \label{fig:model_pwn}
\end{figure*}

We show the obtained SEDs for the three generations of electrons in Fig.~\ref{fig:model_pwn}, together with the observed data.
The model describes the spectra measured with \hess and Suzaku well, and the predicted X-ray flux of the `medium-age' electrons does not exceed the Suzaku upper limit.
The fit yields, for example, a moderate required present-day magnetic field of $\sim$\SI{4}{\micro\gauss} and a reasonable spectral index for the injection spectrum of $\sim$2.
Furthermore, a maximum electron energy of several hundred~TeV is implied by the data.
The total predicted \gam-ray spectrum is also well compatible with the total flux from \hessj as measured by HAWC \citep{Goodman2022}.

The model fails, however, to explain the spectrum of the \fermilat source \fermijten below $\sim$\SI{10}{\GeV}.
This would require an additional IC component, emitted by electrons even older than the `relic' electrons.
In this case, however, it would be expected that the emission of \fermijten exhibits a larger spatial extent than that of component~A of \hessj, which is not the case.
Alternatively, a hadronic component related to the SNR \snrS could be invoked -- this scenario will be discussed in more detail in Section~\ref{sec:hadronic_scenario}.

The offset between component~B and \psrj may be explained, for example, by the proper motion of the pulsar.
Indeed, \citet{Klingler2018,Klingler2020} have detected a northward proper motion of $\sim$\SIrange{20}{40}{\mas\per\year}, albeit not with high significance.
This would imply a travel time between the best-fit position of component~B and the current pulsar position of $\sim$\SIrange{10}{20}{\kilo\year}.
This is somewhat larger than our estimate of the age of the `medium-age' electrons associated with component~B.
However, considering that also an asymmetric crushing of the PWN by the SNR reverse shock can lead to a displacement between the PWN and the pulsar \citep{Blondin2001}, the scenario still appears feasible.

\begin{figure}
  \centering
  \includegraphics{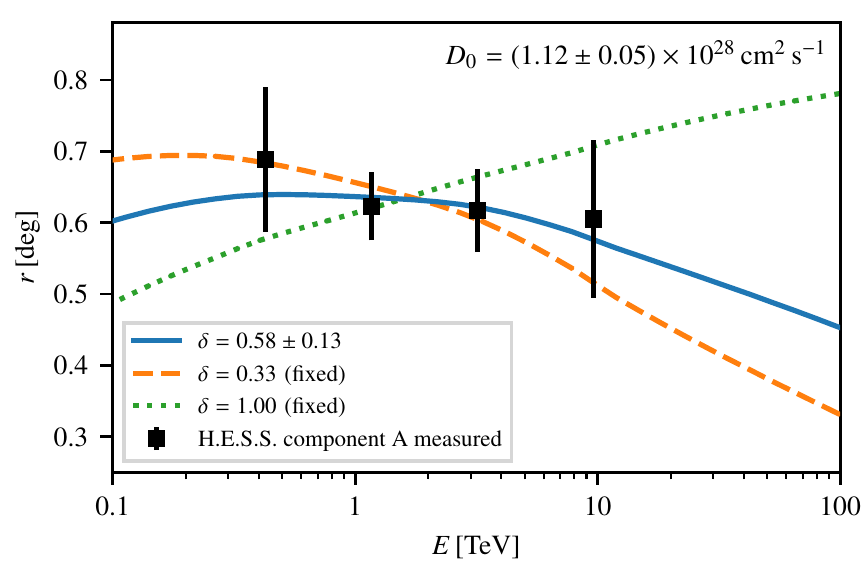}
  \caption{
    Measured and predicted radius of the extended component of \hessj (component~A).
    As measured radius, we used the 1-$\sigma$ extent of the semi-major axis of the elongated Gaussian spatial model for component~A (cf.\ Table~\ref{tab:size_comp1}).
    The solid blue curve has been obtained by fitting the electron diffusion parameters ($D_0, \delta$; cf.\ eq.\ \ref{eq:diff}) to the measured data points; the dashed orange and dotted green curves show results for fixed values of $\delta$ as indicated in the legend.
  }
  \label{fig:hess_comp1_size}
\end{figure}

Having derived the expected age of the PWN system, we used our measurement of the size of the extended \hess component to infer how fast the `relic' electrons associated to this component have diffused since their injection (see Fig.~\ref{fig:hess_comp1_size}).
We have assumed an energy-dependent diffusion coefficient
\begin{linenomath*}
\begin{equation}\label{eq:diff}
  D = D_0\left(\frac{E_e}{\SI{40}{\TeV}}\right)^\delta\,,
\end{equation}
\end{linenomath*}
where $E_e$ is the electron energy, $D_0$ denotes the diffusion coefficient at a reference energy of \SI{40}{\TeV}, and $\delta$ specifies the energy dependence of the diffusion.
Using again the GAMERA library to derive the expected size of the `relic' electron component as a function of \gam-ray energy, we determined the two parameters $D_0$ and $\delta$ by fitting the expected size to the observed size of component~A of \hessj (cf.\ Table~\ref{tab:size_comp1}) -- noting again that the results of the fit are strongly model-dependent and should not be taken as a measurement.
The best-fit diffusion coefficient of $D_0\sim\SI{1.1e28}{\centi\meter\squared\per\second}$ appears reasonable and is of the same order of magnitude as the coefficient obtained for the Geminga halo by \citet{HAWC2017}.
On the other hand, the observed data do not provide very strong constraints for $\delta$, with both Kolgoromov scaling ($\delta=1/3$) and Bohm scaling ($\delta=1$) consistent with the observations.
While our simple estimate assumes a radially symmetric diffusion of the electrons, we note that the elongation of component~A aligns with the asymmetric extension of the X-ray PWN, possibly hinting at a particular arrangement of the magnetic field in the region.
Lastly, we point out that because the highest-energy `relic' electrons have cooled since they were injected, a cut-off to the corresponding \gam-ray spectrum is expected to occur.
The measured cut-off energy of $\sim$\SI{13}{\TeV} for component~A of \hessj is well in line with this prediction, as can be seen in Fig.~\ref{fig:model_pwn_zoom}.

We conclude that the appearance of \hessj is compatible with that of a halo of old electrons (component~A) around the PWN (component~B \& X-ray emission).
We also note that in terms of its X-ray-to-TeV luminosity ratio, \psrj fits well into the population of PWN \citep{Kargaltsev2013}.

\subsection{Possible hadronic contributions}
\label{sec:hadronic_scenario}

Given the presence of SNRs and molecular clouds in the vicinity of \hessj, we also need to consider the possibility that cosmic-ray nuclei accelerated at the SNR shock fronts and interacting hadronically in the molecular clouds are responsible for at least part of the observed \gam-ray emission.
Indeed, a mixed leptonic/hadronic scenario seems possible in principle: while we are not aware of distance estimates for \snrN, existing distance estimates for \snrS of \SI{2.6}{\kpc} \citep{Bamba2003}, $\sim$\SI{3}{\kpc} \citep{Castelletti2016}, and $2.4\pm0.7\,\si{\kpc}$ \citep{Shan2018} seem broadly consistent with those for \psrj of \SI{3.7}{\kpc} \citep{Morris2002} and \SI{3.27}{\kpc} \citep{Parthasarathy2019}.
Furthermore, molecular gas is present throughout the region (cf.\ Appendix~\ref{sec:appendix_fugin_map}), and in particular the dense molecular clouds found by \citet{Castelletti2016} and \citet{Voisin2019} seem to lie at distances compatible with that of \snrS, thus providing the required target material for cosmic-ray interactions.
This has lead \citet{Voisin2019} to propose that \snrS is the host SNR of \psrj.
However, while the pulsar proper motion could be compatible with this scenario, the association is not firm \citep{Klingler2018,Klingler2020}.

Although the measured spectrum of the \fermilat source \fermijten is comparatively soft, it could in principle be described (below $\sim$\SI{10}{GeV}) using a hadronic model.
However, it remains unclear why the spatial model of \fermijten coincides with that of the extended \hess component in this case, as we would rather expect the emission to be more compact and centred on the positions of the molecular clouds.
We also note that simultaneously modelling the emission of \fermijten and either of the two \hess components in a purely hadronic scenario would require, in order to explain the transition from the steep spectrum of the former to the harder spectra of the latter, a spectral hardening in the primary cosmic-ray spectrum, for which there is no obvious explanation.

\begin{figure}
    \centering
    \includegraphics{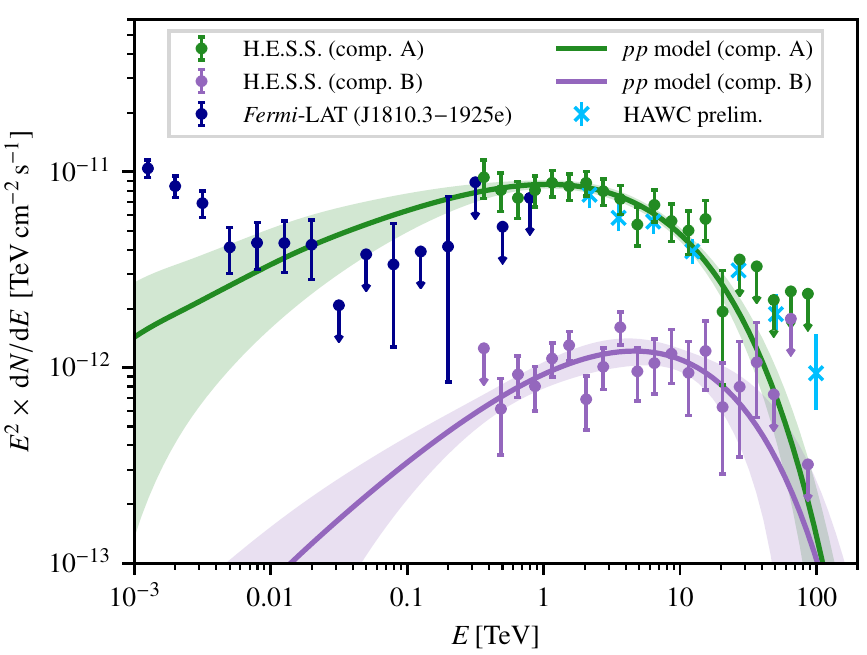}
    \caption{
      SED of \hessj with hadronic ($pp$) models.
      The \hess and \fermilat data points have been derived in this work, the HAWC data points (shown for comparison only) are taken from \citet{Goodman2022}.
      The lines show the predicted \gam-ray spectra for hadronic models fitted to each of the two \hess components, respectively.
    }
    \label{fig:model_snr}
\end{figure}

As presented in Sect.~\ref{sec:pwn_scenario}, both components of \hessj can be modelled well within a leptonic scenario.
Nevertheless, we have also explored the implications of either of the components being of hadronic origin.
To this end, we have fitted a proton-proton ($pp$) model to both components, employing the \textsc{Naima} package \citep{Zabalza2015}.
The primary proton spectrum is described using an ECPL model (see Eq.~\ref{eq:ecpl}) and we have assumed a distance to the source of \SI{3}{\kpc}.
We used the wrapper class for \textsc{Naima} models implemented in \gammapy, so that they could be fitted directly to the \hess data sets (as opposed to fitting them to the extracted flux points only), using the same likelihood framework as before (cf.\ Sect.~\ref{sec:hess_likelihood_analysis}).
The fit results are presented in Table~\ref{tab:pp_model_pars}, and the resulting spectra displayed in Fig.~\ref{fig:model_snr}.
The same spatial models as in the previous analysis (cf.\ Sect.~\ref{sec:hess_results}) were assumed and compatible best-fit parameters were obtained for them.

\begin{table}
  \centering
  \caption{Best-fit parameter values for the hadronic $pp$ models.
  }
  \label{tab:pp_model_pars}
  \begin{tabular}{lc}
    \hline\hline
    Par. [unit] & Value \\\hline
    \multicolumn{2}{c}{\rule{0pt}{1.1\normalbaselineskip} Component~A \vspace{0.05cm}}\\\hline
    $N_0^{p,\mathrm{A}}\,[10^{34}\,\mathrm{eV}^{-1}]$ & $10.0\pm 1.8$\\
    $\Gamma^{p,\mathrm{A}}$ & $1.76\pm 0.17$\\
    $E_c^{p,\mathrm{A}}$ [TeV] & $90_{-30}^{+35}$\\
    $E_0^{p,\mathrm{A}}$ [TeV] & 20 (fixed)\\\hline
    \multicolumn{2}{c}{\rule{0pt}{1.1\normalbaselineskip} Component~B \vspace{0.05cm}}\\\hline
    $N_0^{p,\mathrm{B}}\,[10^{34}\,\mathrm{eV}^{-1}]$ & $1.0\pm 0.4$\\
    $\Gamma^{p,\mathrm{B}}$ & $1.34\pm 0.45$\\
    $E_c^{p,\mathrm{B}}$ [TeV] & $110_{-50}^{+135}$\\
    $E_0^{p,\mathrm{B}}$ [TeV] & 20 (fixed)\\\hline
  \end{tabular}
  \tablefoot{
    $N_0$, $\Gamma$, $E_0$, and $E_c$ are parameters of the ECPL spectral model (Eq.~\ref{eq:ecpl}), where the superscript $p,\mathrm{X}$ denotes that these are the parameters of the primary proton spectrum of component $\mathrm{X}=\{\mathrm{A},\mathrm{B}\}$, respectively.
    The quoted errors represent statistical uncertainties only.
  }
\end{table}

The $pp$ model for component~A prefers a relatively hard spectral index of $\Gamma^{p,\mathrm{A}}=1.76\pm 0.17$.
Integrating the primary spectrum above \SI{1}{\GeV} yields a total required energy of $W^{p,\mathrm{A}}\sim \num{3.2e50}\,(n/\SI{1}{\per\cubic\centi\meter})^{-1}\,\si{\erg}$, which represents -- unless high ISM densities are invoked -- a significant fraction of the canonically assumed kinetic energy released in a supernova explosion of $\sim$\SI{e51}{\erg} \citep[e.g.][]{Ginzburg1975}.
In this context, we note that while the FUGIN CO data indeed show the presence of molecular gas beyond the dense clouds discovered by \citet{Castelletti2016} and \citet{Voisin2019}, they also indicate a gradient in the gas density across the extent of \hessj (see the map in Appendix~\ref{sec:appendix_fugin_map}).
This gradient is not reflected in the observed \gam-ray emission, which makes the interpretation of component~A in a hadronic scenario challenging.

For component~B, our fit yields an even harder proton spectral index of $\Gamma^{p,\mathrm{B}}=1.35\pm 0.45$.
The required energy in protons above \SI{1}{\GeV} is $W^{p,\mathrm{B}}\sim \num{2.7e49}\,(n/\SI{1}{\per\cubic\centi\meter})^{-1}\,\si{\erg}$.
Even considering that only part of the cosmic rays potentially accelerated by \snrS will reach the dense molecular clouds, the high density of gas in the vicinity of component~B in general ($\sim$\SI{1000}{\per\cubic\centi\meter}, cf.\ Appendix~\ref{sec:appendix_fugin_map}) makes this energy seem well affordable.
An explanation of component~B of \hessj in a hadronic scenario therefore appears entirely reasonable, and would furthermore explain its offset from the position of \psrj and the X-ray PWN without the need of requiring, for example, a large proper motion velocity of the pulsar.
It would appear natural in this case to associate component~B also with the highest-energy emission up to \SI{100}{\TeV} measured with HAWC.
However, as is evident from Fig.~\ref{fig:model_snr}, the fitted cut-off energy $E_c^{p,\mathrm{B}}\sim\SI{110}{\TeV}$ of the primary proton spectrum leads to a too strong cut-off in the \gam-ray spectrum, leaving the highest HAWC flux points unexplained.
We have therefore repeated the fit of the hadronic models, adding a $\chi^2$ term that denotes the deviation between the sum of the predicted \gam-ray fluxes of both \hess components and the HAWC flux points to the total TS.
In this case, we obtain slightly softer spectra ($\Gamma^{p,\mathrm{A}}=1.95\pm0.10$; $\Gamma^{p,\mathrm{B}}=1.56\pm0.22$) and slightly higher cut-off energies ($E_c^{p,\mathrm{A}}=140_{-50}^{+80}\,\si{\TeV}$; $E_c^{p,\mathrm{B}}=200_{-130}^{+420}\,\si{\TeV}$).
These values are, however, consistent within uncertainties with those obtained in the previous fit, which demonstrates that it is possible to also explain the measured HAWC flux points within a hadronic scenario (or a mixed one, in which component~A is leptonic and component~B is hadronic).

Finally, we note that while the relatively hard primary proton spectra obtained for both components of \hessj are not consistent with generic predictions of diffuse shock acceleration \citep{Bell2013}, they are compatible with a scenario in which cosmic rays accelerated in a supernova remnant illuminate a nearby gas cloud \citep[e.g.][]{Gabici2009}.
On the other hand, there is also the possibility of a continuous wind of hadronic cosmic rays powered by the pulsar \citep[e.g.][]{Gallant1994,Amato2003}, which may be an interesting scenario to explore for the case of \psrj, as already noted by \citet{Voisin2019}.

\section{Conclusion}
\label{sec:conclusion}

We have presented a new analysis of the \gam-ray emission from \hessj, employing improved analysis techniques.
For the first time, we were able to resolve the emission into two distinct components, which we have modelled with Gaussian spatial models.
Component~A appears extended and elongated, with a 1-$\sigma$ semi-major and semi-minor axis of $\sim0.62^\circ$ and $\sim 0.35^\circ$, respectively, and exhibits a spectrum with a cut-off at $\sim$\SI{13}{\TeV}.
Superimposed, component~B appears symmetric and more compact with a 1-$\sigma$ radius of $\sim0.1^\circ$, and shows a harder spectrum with no clear cut-off.

We have interpreted the results in a leptonic scenario, in which the \gam-ray emission is due to high-energy electrons provided by the energetic pulsar \psrj, which is known to power an X-ray PWN.
The model is based on three `generations' of electrons, associated with component~A, component~B, and the X-ray PWN, respectively (going from old to recently injected electrons).
The measured extent and spectrum of component~A are compatible with a halo of old electrons that have escaped the PWN.

The presence of SNRs and molecular clouds within the region suggests that (part of) the \gam-ray emission could also be of hadronic origin.
Indeed, we found that both of the components of \hessj can in principle be modelled within a hadronic scenario.
However, a lack of correlation between the \gam-ray emission of component~A and the gas present in the region disfavours a hadronic interpretation for this component.
Conversely, for component~B, which is spatially coincident with the shell of the SNR \snrS and several molecular clouds, this is a viable alternative explanation.
The measurement of \gam-ray emission up to \SI{100}{\TeV} with HAWC could be viewed as additional support for this interpretation.
It would, however, leave the X-ray PWN without a counterpart at TeV energies (component~A being associated with electrons injected long ago only), which would be unexpected when comparing with other PWN systems \citep{Kargaltsev2013}.

Our analysis of \fermilat data has confirmed the presence of an extended source, \fermijten, that based on its location and morphology appears to be associated to component~A of \hessj.
However, the spectrum of \fermijten does not connect smoothly to that of component~A of \hessj, implying the need for a spectral hardening around \SI{100}{\GeV}.
While our presented model is not able to describe this feature, we note that the overall shape of the SED is reminiscent of that of another well-known PWN system, Vela~X, which also exhibits a break at around \SI{100}{\GeV} \citep{Tibaldo2018}.
However, multiple distinct components have not been resolved at TeV energies for this system yet.
Furthermore, with its characteristic age of only $\sim$\SI{10}{kyr} \citep{Manchester2005} and a very low braking index of $n=1.4$ \citep{Lyne1996}, the Vela pulsar has an evolution history quite different from \psrj.

Another interesting PWN to compare to is HESS~J1825$-$137, which is the prototype of an extended ($\sim$\SI{100}{\pc} diameter) PWN that shrinks in size at high \gam-ray energies \citep{HESS_J1825_2019}.
The pulsar powering HESS~J1825$-$137, PSR~B1823$-$13 (PSR~J1826$-$1334), is quite similar to \psrj in terms of spin-down power ($\dot{E}=\SI{2.8e36}{\erg\per\second}$), period ($P=\SI{101}{\milli\second}$), and distance ($d=\SI{3.6}{\kpc}$), but may be slightly younger (characteristic age $\tau_c=\SI{21.4}{\kilo\year}$) \citep{Manchester2005}.
Comparing their \gam-ray PWN, \hessj is somewhat less extended than HESS~J1825$-$137 and does not exhibit an energy-dependent morphology.
On the other hand, \hessj seems to be composed of two distinct components, whereas HESS~J1825$-$137 can be modelled with a single component that decreases in extent with increasing energy.
This may suggest that the PWN systems have evolved differently, for example due to differences in the density of the surrounding ISM, or due to a different evolution of the corresponding pulsar (e.g.\ \citeauthor{Khangulyan2018} \citeyear{Khangulyan2018} have proposed an unusually short birth period of $P_0\sim\SI{1}{\milli\second}$ for PSR~B1823$-$13).

Finally, it is interesting that \hessj shows characteristics very similar to those of HESS~J1702$-$420 \citep{HESS_J1702_2021}: both have been resolved into a compact, hard-spectrum component surrounded by an extended, softer-spectrum component.
This may in principle suggest a similar origin of the \gam-ray emission, however, HESS~J1702$-$420 is a `dark' source that lacks an obvious counterpart at other wavelengths \citep[see also][]{Giunti2022}, hampering a further comparison with \hessj.

While we are not able to draw definitive conclusions about the origin of the \gam-ray emission of \hessj, our detailed and simultaneous characterisation of its morphology and spectrum is a big step towards understanding this source.
Further observations, in particular with HAWC \citep{HAWC2017a} as well as with the upcoming Cherenkov Telescope Array \citep[CTA;][]{CTA2018} and Southern Wide-Field Gamma-Ray Observatory \citep[SWGO;][]{SWGO2019}, will be crucial in further broadening our knowledge about \hessj.

\begin{acknowledgements}
The support of the Namibian authorities and of the University of Namibia in facilitating the construction and operation of \hess is gratefully acknowledged, as is the support by
the German Ministry for Education and Research (BMBF),
the Max Planck Society,
the German Research Foundation (DFG),
the Helmholtz Association,
the Alexander von Humboldt Foundation,
the French Ministry of Higher Education, Research and Innovation, 
the Centre National de la Recherche Scientifique (CNRS/IN2P3 and CNRS/INSU),
the Commissariat \`{a} l'\'{E}nergie atomique et aux \'{E}nergies alternatives (CEA), 
the U.K. Science and Technology Facilities Council (STFC),
the Irish Research Council (IRC) and the Science Foundation Ireland (SFI),
the Knut and Alice Wallenberg Foundation, 
the Polish Ministry of Education and Science, agreement no.~2021/WK/06,
the South African Department of Science and Technology and National Research Foundation, 
the University of Namibia,
the National Commission on Research, Science \& Technology of Namibia (NCRST),
the Austrian Federal Ministry of Education, Science and Research and the Austrian Science Fund (FWF),
the Australian Research Council (ARC),
the Japan Society for the Promotion of Science,
the University of Amsterdam and
the Science Committee of Armenia grant 21AG-1C085.
We appreciate the excellent work of the technical support staff in Berlin, Zeuthen, Heidelberg, Palaiseau, Paris, Saclay, T\"{u}bingen and in Namibia in the construction and operation of the equipment.
This work benefited from services provided by the \hess Virtual Organisation, supported by the national resource providers of the EGI Federation.
This research made use of the
\textsc{Astropy}\footnote{\url{https://www.astropy.org}} \citep{Robitaille2013,PriceWhelan2018},
\textsc{Matplotlib}\footnote{\url{https://matplotlib.org}} \citep{Hunter2007},
and
\textsc{Corner}\footnote{\url{https://corner.readthedocs.io}} \citep{ForemanMackey2016}
software packages.
\end{acknowledgements}

\bibliographystyle{aa}
\bibliography{references}

\begin{appendix}

\section{Fit of hadronic background model}
\label{sec:appendix_bkg_fit}

We describe in this appendix the fit of the hadronic background model to the analysed observation runs.
This procedure is necessary to ensure a valid description of the background for all runs.

\begin{figure}[ht]
  \centering
  \includegraphics{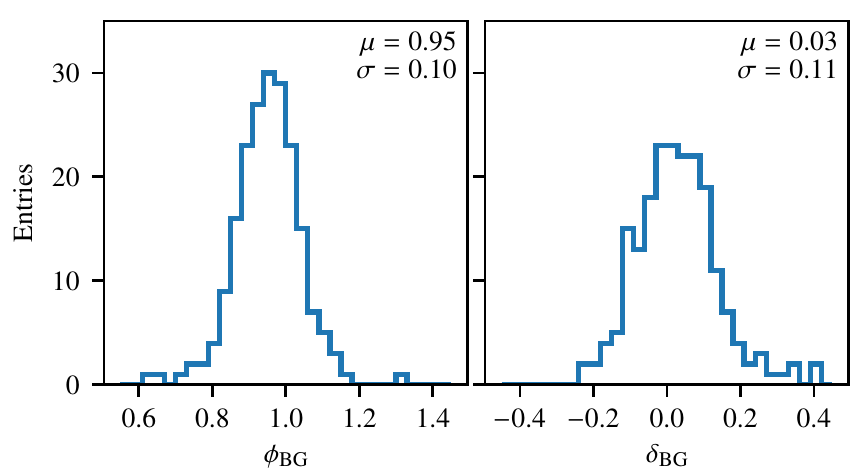}
  \caption{
    Distributions of fitted background model parameters $\phi_\mathrm{BG}$ and $\delta_\mathrm{BG}$.
    The mean ($\mu$) and standard deviation ($\sigma$) of the distributions are specified in the panels.
  }
  \label{fig:bkg_fit_pars}
\end{figure}

\begin{figure}[ht]
  \centering
  \includegraphics{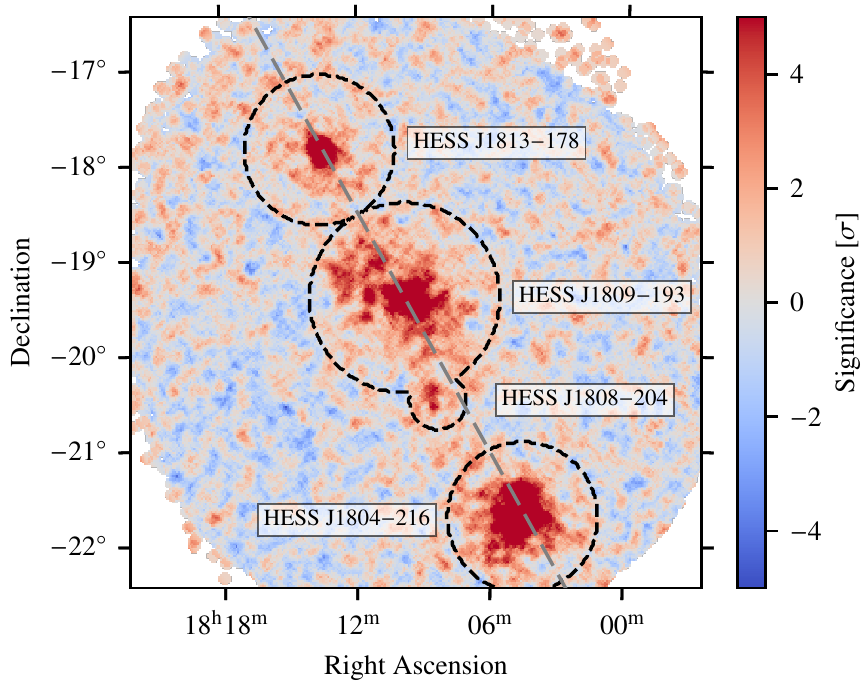}
  \caption{
    Significance map after fit of hadronic background model.
    We used an oversampling radius of $0.07^\circ$ for smoothing.
    Besides \hessj, the $6^\circ\times 6^\circ$ RoI contains the \hess sources HESS~J1804$-$216, HESS~J1808$-$204 and HESS~J1813$-$178.
    The grey dashed line marks the Galactic plane, and the black dashed circles show exclusion regions as used in the fit of the hadronic background model.
  }
  \label{fig:sign_map_bkg_fit}
\end{figure}

We adjusted the background model by running the 3D likelihood fit without any source components included, whereby we excluded regions around known \gam-ray sources from the fit (as indicated in Fig.~\ref{fig:sign_map_bkg_fit}).
For each observation, we fitted the overall background normalisation ($\phi_\mathrm{BG}$) and a spectral tilt parameter ($\delta_\mathrm{BG}$), which modifies the predicted background rate $R_\mathrm{BG}$ at energy $E$ as
\begin{linenomath*}
\begin{equation}\label{eq:bkg_delta}
    R^{\prime}_\mathrm{BG}=R_\mathrm{BG}\cdot(E/E_0)^{-\delta_\mathrm{BG}}\,,
\end{equation}
\end{linenomath*}
with reference energy $E_0=1\,\mathrm{TeV}$, to correct for small inaccuracies of the spectral shape of the background model.
Figure~\ref{fig:bkg_fit_pars} displays distributions of the fitted parameter values.
In Fig.~\ref{fig:sign_map_bkg_fit}, we show the resulting significance map obtained with all observations, which we have smoothed using a top hat kernel with $0.07^\circ$ radius, which corresponds approximately to the size of the point spread function of \hess for this analysis.
We computed the significance for each spatial pixel following \citet{Li1983}, assuming a negligible uncertainty of the predicted number of background events, which is justified when considering the entire energy range of the analysis, as we did here.
This is also known as the `Cash' statistic \citep[see][]{Cash1979}.

\begin{figure}[ht]
  \centering
  \includegraphics{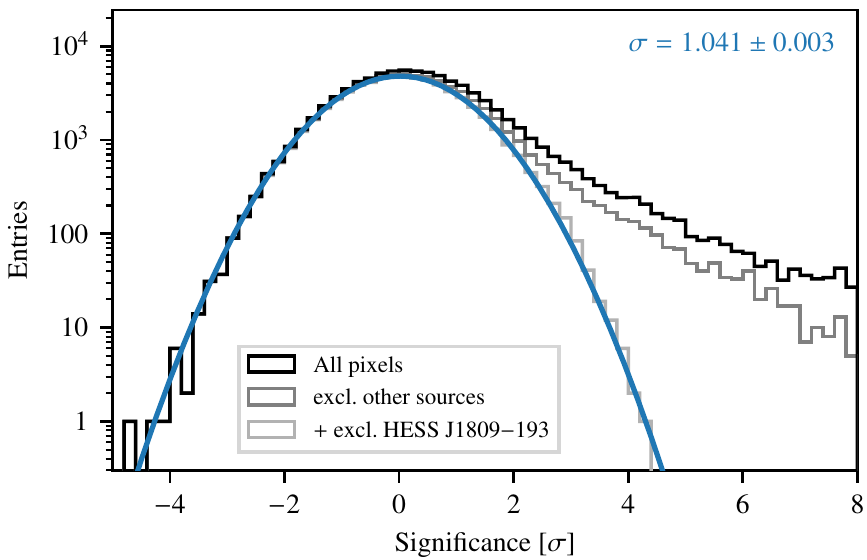}
  \caption{
    Significance distributions for hadronic background model fit.
    The black histogram shows the distribution of significance values of all spatial pixels of the map in Fig.~\ref{fig:sign_map_bkg_fit}.
    The dark grey histogram is for all pixels outside the exclusion regions around HESS~J1804$-$216, HESS~J1808$-$204, and HESS~J1813$-$178 (i.e.\ contains only the excess from \hessj), and the light grey histogram is for all pixels outside any of the exclusion regions.
    The blue line shows the result of fitting a Gaussian distribution to the light grey histogram, the fitted width of the Gaussian is given in the upper right corner of the plot.
  }
  \label{fig:sign_dist}
\end{figure}

While significant \gam-ray emission is clearly visible in each of the exclusion regions, no significant deviation from the predicted background is present outside these regions.
This picture is confirmed by the distribution of significance values in all map pixels, which we show in Fig.~\ref{fig:sign_dist}.
That the distribution for all pixels outside the exclusion regions closely follows a Gaussian distribution with unity width -- the expectation for purely statistical fluctuations -- indicates that we have achieved a very good description of the hadronic background.
Assuming the excess in width above unity is due to a systematic effect that scales the background rate by a constant factor, the observed width of $\sigma=1.041$ implies a level of background systematics of $\sim$2\% for the studied data set.

\clearpage

\section{Estimation of systematic uncertainties}
\label{sec:appendix_sys_err}

We list in Table~\ref{tab:sys_err_pars} the parameters that we have used to vary the IRFs in order to estimate systematic uncertainties for our fit results.
The variation of the global energy scale is implemented such that the energy axes of the effective area, energy dispersion, and point spread function IRFs are scaled by a factor $\phi_E$, simulating the effect of a wrongly calibrated energy scale.
We have sampled the applied factors $\phi_E$ from a Gaussian distribution with 10\% width, which approximately corresponds to a shift of the energy scale as may be expected, for example, from atmospheric variations \citep{Hahn2014}.

For the background models, we have considered three different systematic variations: besides the overall normalisation ($\phi_\mathrm{BG}$) and spectral tilt parameter ($\delta_\mathrm{BG}$) that were already introduced in Appendix~\ref{sec:appendix_bkg_fit}, we have applied in addition a linear gradient across the field of view, with amplitude $A^\mathrm{grad}_\mathrm{BG}$ and direction angle $\alpha^\mathrm{grad}_\mathrm{BG}$.
To choose the magnitude of the variation, we have compared the (already adjusted) background model to the observed data.
The adopted values reflect the variation of the parameters that may still reasonably be expected for the entire data set after the adjustment described in Appendix~\ref{sec:appendix_bkg_fit}.
A gradient of the background in the field of view may, for example, arise due to unmodelled diffuse \gam-ray emission. 
All systematic parameters are assumed to affect the entire data set (i.e.\ all observation runs) in the same way.

The procedure to derive the systematic errors is then the following:
\begin{enumerate}[(a)]
    \item Sample random values for the variation parameters from the distributions indicated in Table~\ref{tab:sys_err_pars}.
    \item Apply systematic variations to the IRFs of all data sets.
    \item Generate pseudo data sets with randomised observed counts by sampling for each bin from a Poisson distribution with mean equal to the number of events predicted by the background model and best-fit source models, using the varied IRFs to compute the latter.
    \item Re-perform likelihood fit, using the original, unmodified IRFs of the data sets.
    \item Repeat steps (a)--(d) $2500$ times.
    \item Determine the spread of the distributions of fitted source parameters, and compute systematic errors by subtracting quadratically their statistical uncertainties.
\end{enumerate}

\begin{figure}[ht]
  \centering
  \subfigure[]{
    \includegraphics{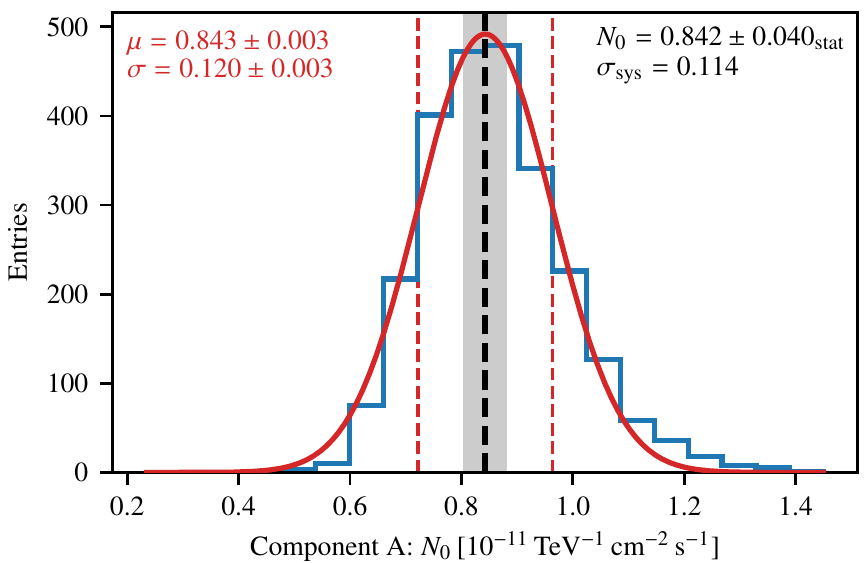}
    \label{fig:sys_err_amp_comp1}
  }
  \subfigure[]{
    \includegraphics{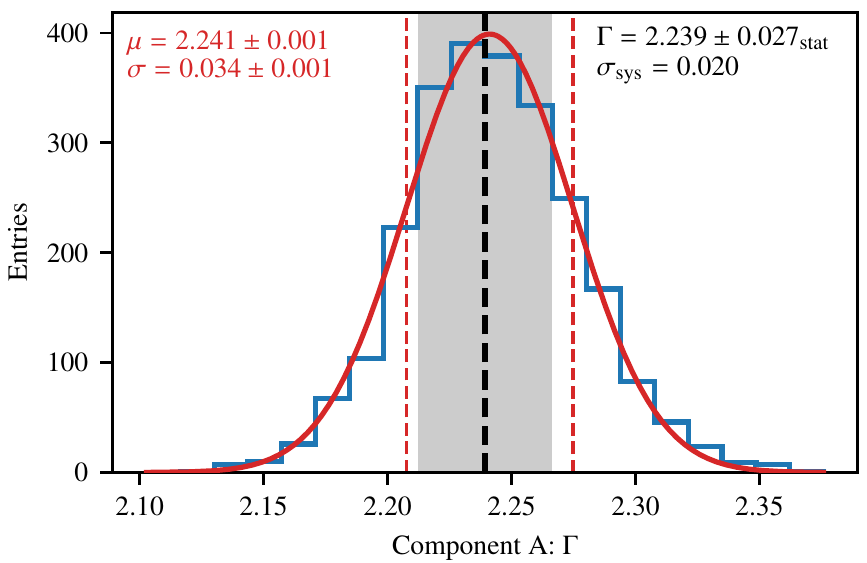}
    \label{fig:sys_err_index_comp1}
  }
  \caption{Source parameter distributions for two example parameters.
  (a) Flux normalisation $N_0$ of source component~A.
  (b) Power-law index $\Gamma$ of source component~A.
  The black dashed line and grey band denote the best-fit value and statistical uncertainty of this parameter, also specified in the top right corner.
  The red line shows the fit of a Gaussian to the histogram, where the fitted mean and width are indicated in the top left corner.
  The red dashed lines indicate the 1-$\sigma$ width of the fitted Gaussian, which represents the total (i.e.\ statistical and systematic) uncertainty on this parameter.
  The derived systematic error, obtained by subtracting quadratically the statistical error from the fitted total width, is also stated in the figure.
  }
  \label{fig:sys_err_par_dists}
\end{figure}

Step (f) of the above procedure is illustrated in Fig.~\ref{fig:sys_err_par_dists}, where we show two example parameter distributions, corresponding to the amplitude $N_0$ and spectral index $\Gamma$ of source model component~A (cf.\ Sect.~\ref{sec:hess_results}).
For both parameters, the spread of the distribution (indicated by the red dashed lines) exceeds the expectation from purely statistical fluctuations (grey band), which demonstrates that the considered systematic effects affect the uncertainties of these parameters.

\begin{figure*}
    \centering
    \includegraphics{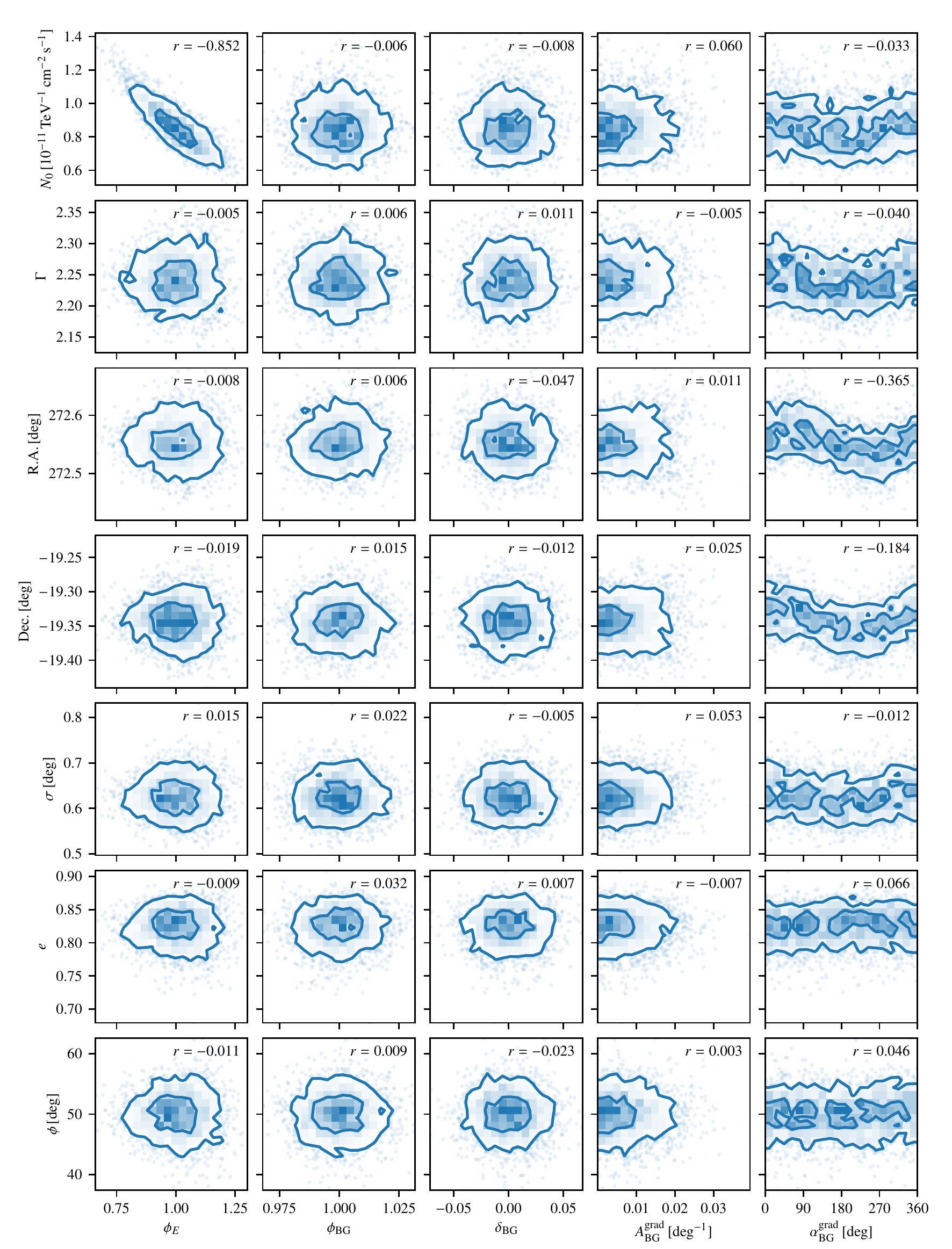}
    \caption{
      Correlation plots between the source model parameters of source component~A (rows) and the systematic variation parameters (columns).
      The value specified in the top right corner of each panel is Pearson's correlation coefficient.
    }
    \label{fig:sys_err_scatter_comp1}
\end{figure*}

Finally, we illustrate the correlation between the source model parameters (again exemplary those of source component~A, cf.\ Sect.~\ref{sec:hess_results}) and the systematic variation parameters in Fig.~\ref{fig:sys_err_scatter_comp1}.\footnote{
  In the figure, we show the absolute value of $A^\mathrm{grad}_\mathrm{BG}$, and have flipped $\alpha^\mathrm{grad}_\mathrm{BG}$ by $180^\circ$ if $A^\mathrm{grad}_\mathrm{BG}$ was negative.
}
As intuitively makes sense, the strongest correlation is observed between the energy scale shift parameter $\phi_E$ and the source model normalisation $N_0$.
The parameters $\phi_\mathrm{BG}$ and $\delta_\mathrm{BG}$ are -- as in our standard fit procedure -- re-adjusted in the fit of the pseudo data sets, hence no strong correlation with the source model parameters is expected.
The parameters of the spatial part of the source model (e.g.\ its R.A.\ and Dec.\ coordinate) are affected mostly by the assumed background model gradient, and a modulation of these parameters with the angle $\alpha^\mathrm{grad}_\mathrm{BG}$ is observed.

While the correlations between the systematic variation parameters and the source model parameters are sometimes weak, they do lead to a slight broadening of the source model parameter distributions in most cases (see e.g.\ the distribution of the spectral index $\Gamma$ shown in Fig.~\ref{fig:sys_err_index_comp1}), and thus enable the estimation of a systematic uncertainty.
We specify the resulting systematic uncertainties for all parameters along with the general fit results in Sect.~\ref{sec:hess_results}, Table~\ref{tab:model_pars}.
For parameters for which the distribution was not broadened by the systematic effects considered here, we specify no systematic error -- this is the case for the eccentricity ($e$) and position angle ($\phi$) parameters of source component~A, as well as for the fitted R.A.\ and Dec.\ coordinate of source component~B.
However, we note that these parameters may be affected by other systematic effects neglected here.
In particular, the fitted source positions are subject to a systematic uncertainty of the pointing position of the \hess telescopes, which is of the order of $10''-20''$ \citep{Gillessen2004}.

\begin{table}[ht]
  \centering
  \caption{Parameter variations for systematic uncertainty estimation.
  }
  \label{tab:sys_err_pars}
  \begin{tabular}{ccc}
    \hline\hline
    Par. & Variation & Description \\\hline
    \multicolumn{3}{c}{\rule{0pt}{1.1\normalbaselineskip} Global energy scale \vspace{0.05cm}}\\\hline
    $\phi_E$ & \makecell{Gaussian\\($\mu=1,\,\sigma=0.1$)} & Shift of energy scale\\\hline
    \multicolumn{3}{c}{\rule{0pt}{1.1\normalbaselineskip} Background model variations \vspace{0.05cm}}\\\hline
    $\phi_\mathrm{BG}$ & \makecell{Gaussian\\($\mu=1,\,\sigma=0.01$)} & \makecell{Background model\\normalisation}\\
    $\delta_\mathrm{BG}$ & \makecell{Gaussian\\($\mu=0,\,\sigma=0.02$)} & \makecell{Background model\\spectral tilt}\\
    $A^\mathrm{grad}_\mathrm{BG}$ & \makecell{Gaussian\\($\mu=1,\,\sigma=0.01$)} & \makecell{Amplitude of background\\gradient (in deg$^{-1}$)}\\
    $\alpha^\mathrm{grad}_\mathrm{BG}$ & \makecell{Uniform\\($0^\circ-360^\circ$)} & \makecell{Direction of background\\gradient}\\\hline
  \end{tabular}
  \tablefoot{For a detailed explanation of the parameters, see text.}
\end{table}

\clearpage

\section{Significance maps in energy bands}
\label{sec:appendix_maps_ebands}

To check the agreement between a fitted model and the observed data, it can be illustrative to investigate residual significance maps in separate energy ranges.
Furthermore, it is possible to search for indications of energy-dependent source morphology by fitting a source model within restricted energy ranges, and observing how the fitted source parameters vary between the different ranges.
For these purposes, we have defined four mutually exclusive energy bands, with lower boundaries at \SI{0.27}{\TeV}, \SI{0.75}{\TeV}, \SI{2.1}{\TeV}, and \SI{5.6}{\TeV}.

\subsection{Fit of 1-component model in energy bands}
\label{sec:appendix_1comp_fit_ebands}

\begin{figure*}[hb]
  \sidecaption
  \includegraphics{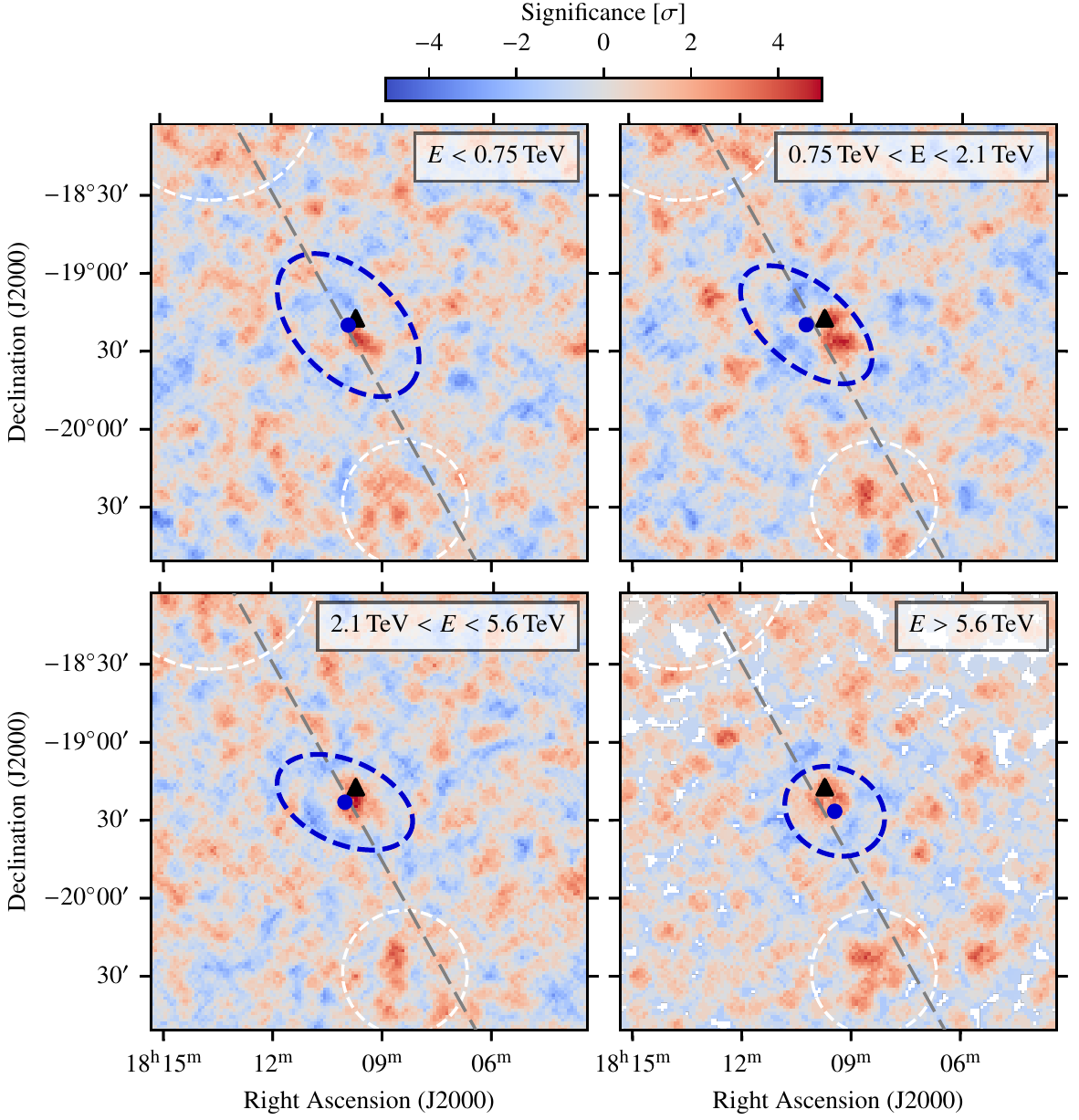}
  \caption{
    Residual significance maps for the 1-component model, fitted in energy bands.
    We used an oversampling radius of $0.07^\circ$ for smoothing.
    Energy bands are specified in the upper right corner of each panel.
    The blue circle markers and dashed lines display the best-fit position and 1-sigma extent of the Gaussian models.
    The grey dashed line marks the Galactic plane, white dashed circles shows regions excluded from the analysis, and the black triangle marker denotes the position of \psrj.
  }
  \label{fig:sign_maps_1comp_ebands}
\end{figure*}

As detailed in the main part of the paper, the 1-component model does not yield a satisfactory description of the observed \hess data.
This has led us to adopt the 2-component model, which provides a much better fit.
We investigate in this section the possibility that a similarly good fit can be obtained by allowing the parameters of the Gaussian spatial model of the 1-component model to vary with energy.
To this end, we have repeated the likelihood fit in the four different energy bands defined above, whereby we kept the spectral index $\Gamma$ fixed to its best-fit value from the fit across all energies ($\Gamma=2.184$).
In Fig.~\ref{fig:sign_maps_1comp_ebands}, we show the resulting residual significance maps for all four energy bands, with the best-fit position and 1-$\sigma$ radius of the fitted component overlaid.
The maps demonstrate that even when the parameters of the spatial model are allowed to vary with energy, the 1-component model cannot describe the data well -- this is most evident from the energy band between \SI{0.75}{\TeV} and \SI{2.1}{\TeV}, where strong residuals are still visible.

\subsection{Residual significance maps for 2-component model in energy bands}
\label{sec:appendix_2comp_maps_ebands}

\begin{figure*}[hb]
  \sidecaption
  \includegraphics{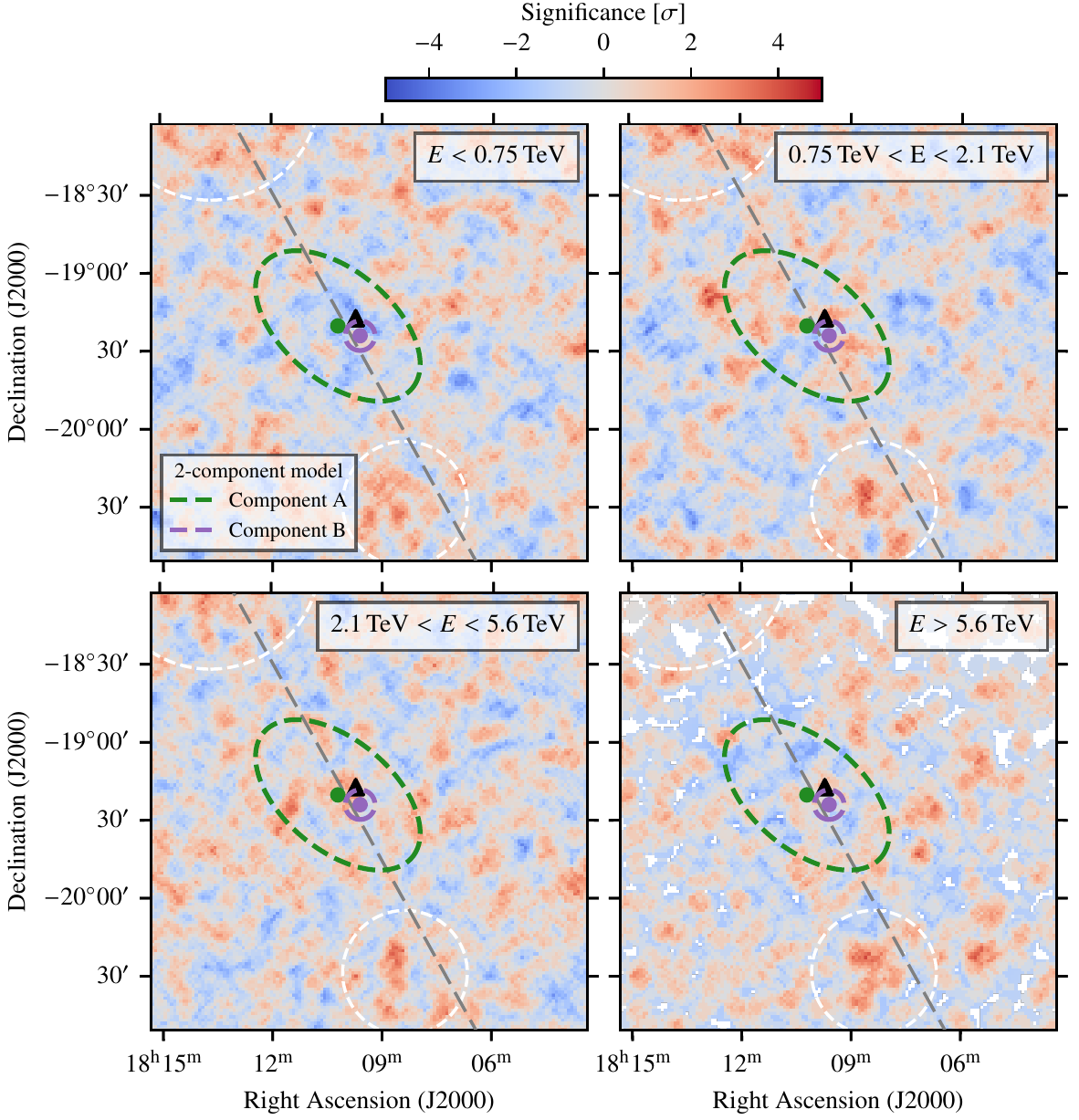}
  \caption{
    Residual significance maps for the 2-component model.
    We used an oversampling radius of $0.07^\circ$ for smoothing.
    Energy bands are specified in the upper right corner of each panel.
    The green and purple circle marker and dashed line display the best-fit position and 1-sigma extent of the Gaussian models of component~A and~B, respectively.
    The grey dashed line marks the Galactic plane, white dashed circles shows regions excluded from the analysis, and the black triangle marker denotes the position of \psrj.
  }
  \label{fig:sign_maps_2comp_fullenergy_ebands}
\end{figure*}

We provide in Fig.~\ref{fig:sign_maps_2comp_fullenergy_ebands} residual significance maps for the 2-component model, computed for the four different energy bands defined above.
Note that these maps have been computed based on the 2-component model fitted across the full energy range (i.e.\ between \SI{0.27}{\TeV} and \SI{100}{\TeV}).
The absence of strong residuals in any of the bands indicates a good agreement between the fitted model and the observed data, at all energies.

\subsection{Fit of 2-component model in energy bands}
\label{sec:appendix_2comp_fit_ebands}

\begin{figure*}[hb]
  \sidecaption
  \includegraphics{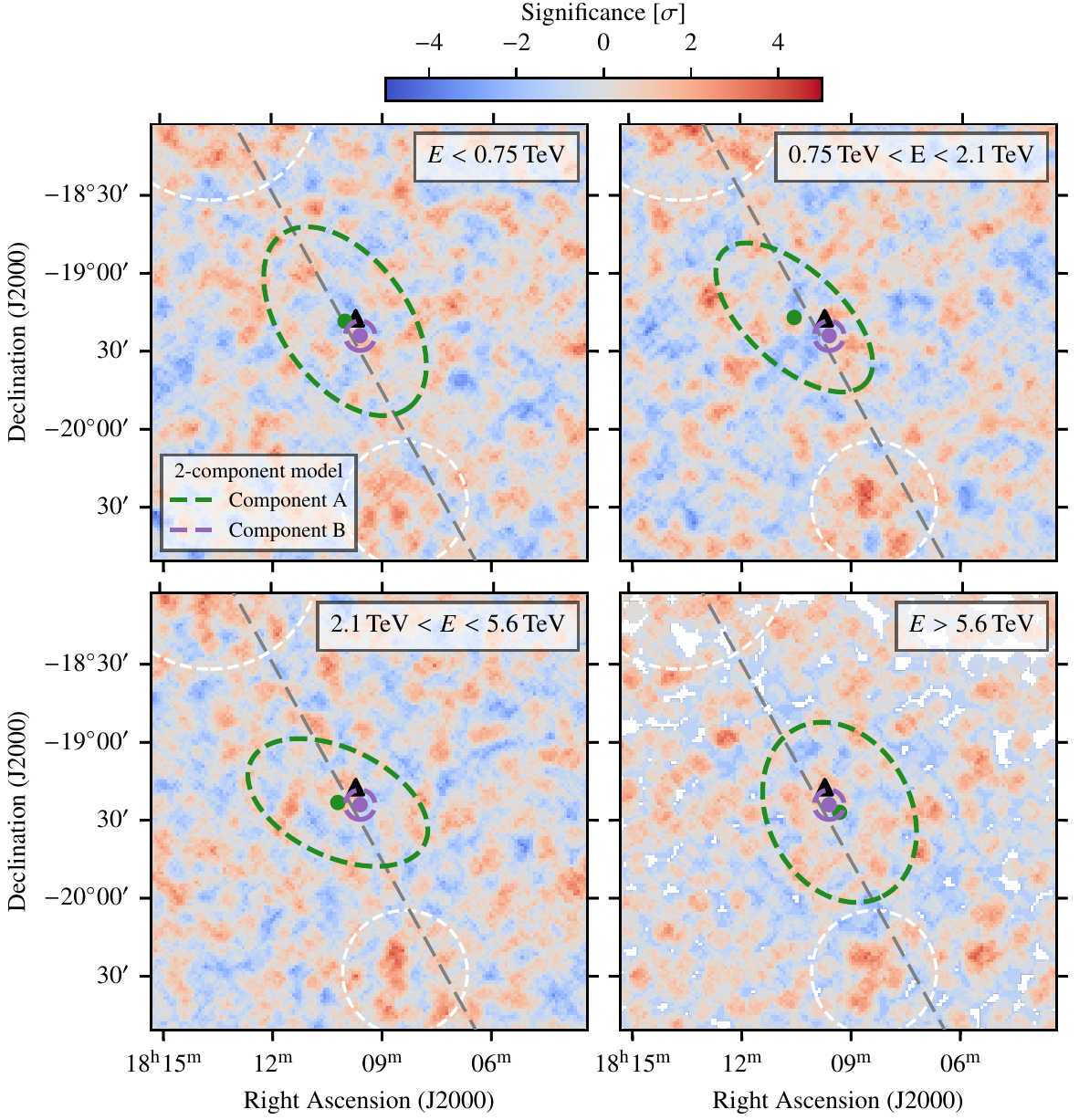}
  \caption{
    Residual significance maps for the 2-component model, fitted in energy bands.
    We used an oversampling radius of $0.07^\circ$ for smoothing.
    Energy bands are specified in the upper right corner of each panel.
    The green and purple circle marker and dashed line display the best-fit position and 1-sigma extent of the Gaussian models of component~A and~B, respectively.
    The grey dashed line marks the Galactic plane, white dashed circles shows regions excluded from the analysis, and the black triangle marker denotes the position of \psrj.
  }
  \label{fig:sign_maps_2comp_ebands}
\end{figure*}

While the 2-component model yields a satisfactory description of the data across the entire energy range (cf.\ previous section), we also checked for a possible energy-dependent morphology of component~A, by repeating the fit in the same energy bands as above.
In these fits, we fixed the parameters of component~B, as well as the spectral index of component~A.
The resulting residual significance maps are displayed in Fig.~\ref{fig:sign_maps_2comp_ebands}.
While the centre position and shape of the elongated Gaussian model vary slightly between the different energy ranges, those changes are well within the statistical uncertainties.
In particular, as summarised in Table~\ref{tab:size_comp1} in the main text, the fitted extent of component~A shows no significant variation with energy.

\clearpage

\section{FUGIN CO map}
\label{sec:appendix_fugin_map}

\begin{figure}[hb]
  \centering
  \includegraphics{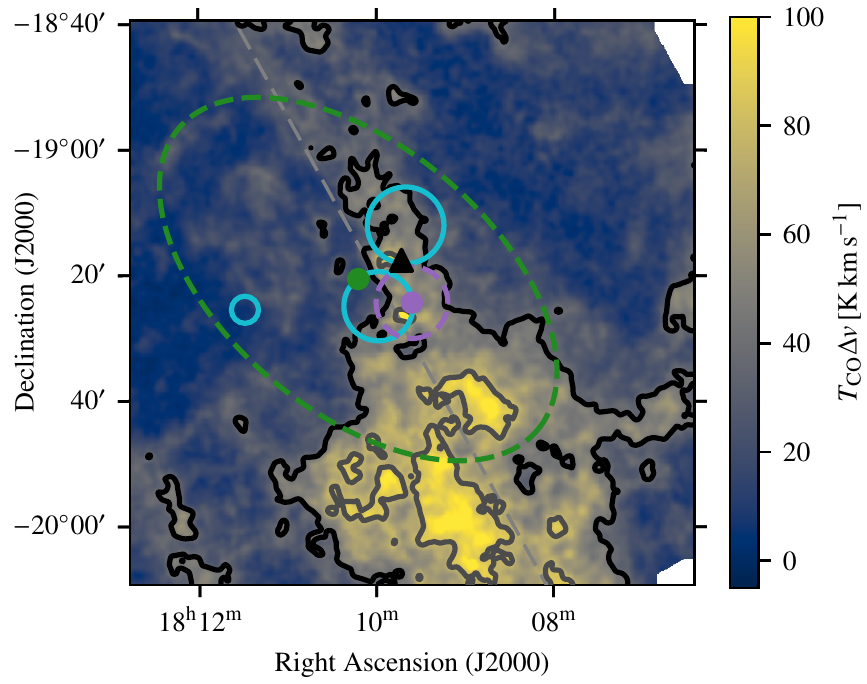}
  \caption{
    Map of FUGIN $^{12}$CO ($J$=1--0) line emission in the region surrounding \psrj \citep{Umemoto2017}.
    Contour lines at \SI{40}{\kelvin\kilo\meter\per\second} and \SI{80}{\kelvin\kilo\meter\per\second} are shown in black and dark grey, respectively.
    The lines at \SI{40}{\kelvin\kilo\meter\per\second} are also displayed in Fig.~\ref{fig:flux_map}(b).
    The black triangle marker denotes the position of \psrj, light blue circles show the positions of SNRs, and the grey dashed line marks the Galactic plane.
    The position and 1-$\sigma$ extent of component~A and component~B of \hessj are displayed in green and purple, respectively.
  }
  \label{fig:fugin_co_map}
\end{figure}

We show in Fig.~\ref{fig:fugin_co_map} a map of the $^{12}$CO~($J$=1--0) emission, which traces molecular hydrogen gas, taken from the FUGIN survey \citep{Umemoto2017}.
The emission is shown for an interval in velocity with respect to the local standard of rest of $v=\SIrange{16}{27}{\kilo\meter\per\second}$ \citep[following][]{Castelletti2016}, which corresponds to a distance of $\sim$\SI{3}{\kpc}.
The map has been smoothed with a Gaussian kernel of $0.5\arcmin{}$ width.
Using an CO-to-H$_2$ conversion factor of $1.5\times 10^{20}\,\mathrm{cm}^{-2}\,/\,(\mathrm{K}\,\mathrm{km}\,\mathrm{s}^{-1})$, and accounting for an additional 20\% of He, we obtained a gas density of $\sim$\SI{1000}{\per\cubic\centi\meter} in the vicinity of component~B of \hessj.

\end{appendix}

\end{document}